\pdfoutput=1
\documentclass[lettersize,journal]{IEEEtran}
\usepackage{longtable}
\usepackage{array}
\usepackage[table,xcdraw]{xcolor}
\usepackage{colortbl}
\usepackage{pdflscape}
\usepackage{adjustbox}
\usepackage{enumitem}
\usepackage{booktabs}
\usepackage{amssymb}
\usepackage{rotating}
\usepackage{graphicx}
\usepackage{listings}
\usepackage{tabularx}
\usepackage{microtype}
\usepackage{verbatim}
\usepackage[utf8]{inputenc}
\usepackage{tcolorbox} 
\usepackage{enumitem}  
\usepackage{multirow}
\usepackage{subcaption}
\usepackage{makecell}
\usepackage{listings}
\usepackage{xcolor}
\usepackage{xurl}
\usepackage{url}
\usepackage[hidelinks]{hyperref}
\usepackage{cleveref}
\AtBeginDocument{}

\newcommand{\tool}{\textit{IssueSupport}}
\newcommand{\toolcapital}{ISSUESUPPORT}

\newcommand{\invalidTotalCount}{8,289}
\newcommand{\filteredInvalidCount}{4,463}
\newcommand{\invalidReportCount}{1,404}
\newcommand{\invalidSampleCount}{302}
\newcommand{\thresholdYear}{2022}

\newcommand{\unchecked}{}

\begin{document}

\title{Automated Root-Cause Subclassification and No-Code Fix Generation for Invalid Bug Reports}

\author{Mahmut Furkan Gön\textsuperscript{*}, 
        Emre Dinç\textsuperscript{*}, 
        Tevfik Emre Sungur\textsuperscript{*}, 
        and Eray Tüzün%
\thanks{\textsuperscript{*}These authors contributed equally to this work.}%
\thanks{All authors are with the Department of Computer Engineering, Bilkent University, Ankara, Türkiye. 
Emails: \{furkan.gon, emre.sungur,\}@bilkent.edu.tr,
eraytuzun@cs.bilkent.edu.tr, emre.dinc@ug.bilkent.edu.tr}}

\markboth{IEEE Transactions on Software Engineering}%
{Gön \MakeLowercase{\textit{et al.}}: \toolcapital: Detecting Invalid Bug Reports}

\maketitle

\begin{abstract}
\textbf{Context:}
Issues faced when using software are reported in the form of bug reports. However, many bug reports are invalid, meaning they do not require code changes and are resolved with a no-code fix. Manually determining the root cause of the invalid bug reports and providing actionable resolutions by the customer support causes a serious waste of resources. 
 
\smallskip
 
\textbf{Objective:} Our goal is to introduce a standardized taxonomy for root-cause oriented invalid bug report subclassification, and perform experiments to test the accuracy of various approaches on invalid subclassification based on root causes and no-code fix generation. We study how different configurations perform on a manually curated and adjudicated evaluation benchmark we have created.
 
\smallskip
 
\textbf{Method:} Using a manually curated benchmark for higher quality analysis, we experimented with vanilla large language models (LLMs), Retrieval Augmented Generation (RAG), and agentic web search to identify invalid subclasses and generate no-code fixes. We evaluated the results against manually labeled ground truth data that includes the invalid subclass and no-code fixes from the original bug reports. We measured subclass detection performance with weighted F1-Score, and assessed no-code fix suggestions using BERTScore and Judge LLM success rates.
 
\smallskip
 
\textbf{Results:} For subclassification, RAG achieves the highest overall performance with 0.66 weighted F1, slightly outperforming vanilla LLMs at 0.65 and agentic web search at 0.64. At the subclass level, performance peaks at 0.85 F1 for Non-reproducibility in the RAG pipeline with Gemini and 0.81 for Feature Request in Vanilla LLM, while Wrong Version remains the most challenging with scores between 0.00 and 0.29. For no-code fix generation, agentic web search with Gemini 3.1 Pro achieves the highest overall Judge LLM success rate at 68.9\%, compared to 64.4\% for RAG and 63.1\% for the same model's vanilla baseline, while the strongest vanilla model overall (GPT-5.4) reaches 64.9\%. Among these pipelines, subclass-level peaks are 88.2\% for Working as Designed under RAG and 74.4\% for Question under agentic web search.
 
\smallskip
 
\textbf{Conclusion:} This study shows that context engineering has task- and model-dependent effects that are often small in magnitude: retrieval augmented generation yields modest gains for invalid subclassification, while no-code fix generation benefits more from agentic web search, indicating that different components are effective for different tasks. The generated no-code fixes demonstrate the potential to automate customer support workflows, reduce manual triage effort, and provide immediate, actionable guidance to users within bug tracking systems.
\end{abstract}


\section{Introduction} \label{sec: intro}


Software maintenance is an essential aspect of software development. Software maintenance costs more than two trillion USD annually \cite{CISQ2022}.  Most open-source and proprietary projects address software maintenance through bug reports in bug tracking platforms such as Jira\footnote{\url{https://www.atlassian.com/software/jira}} or GitHub Issues\footnote{\url{https://github.com/features/issues}}. End users create bug reports that reflect their problems, such as implementation issues, crashes, configuration problems, or misunderstandings about the software product. These reports are then reviewed by project owners or developers, who investigate and resolve the reported problems. From the software maintainers' perspective, bug reports can be categorized as either valid bug reports that require modifications to the source code or invalid bug reports that do not require changes to the source code, hence not inherently a software defect \cite{fan2020, laiq2023, anvik2006should}.

A considerable portion of submitted bug reports are labeled as invalid. For example, in the Eclipse project, about \textbf{22\%} of the 121,855 reported bugs were classified as invalid \cite{fan2020}. While invalid bug reports usually do not require changes to the source code, they may still highlight issues within the product. In an analysis of 7,000 bug reports across five open-source projects, \textbf{33.8\%} of reports initially labeled as valid bugs were actually misclassified \cite{herzig2013}. As a result, \textbf{39\%} of source files marked as defective were found to have never contained an actual bug \cite{herzig2013}. Moreover, one study found that resolving invalid bug reports can take nearly as long as resolving valid ones, \textbf{16 to 18 days} on average across two different software products \cite{laiq2022_ttr}.

There might be several root causes why a bug report can be invalid, including a misunderstanding of an intended feature \cite{laiq2023, sun2011bug, su2017creating, panichella2021won}, user-side configuration problems \cite{laiq2023, sun2011bug, su2017creating, panichella2021won}, or using a wrong version of the software product \cite{laiq2023, su2017creating, panichella2021won}. The customer support team typically identifies the root causes of invalid bug reports \cite{torun2025past}. 

\begin{figure}[t]
    \centering
    
    \begin{subfigure}{\linewidth}
        \centering
        \includegraphics[width=\linewidth]{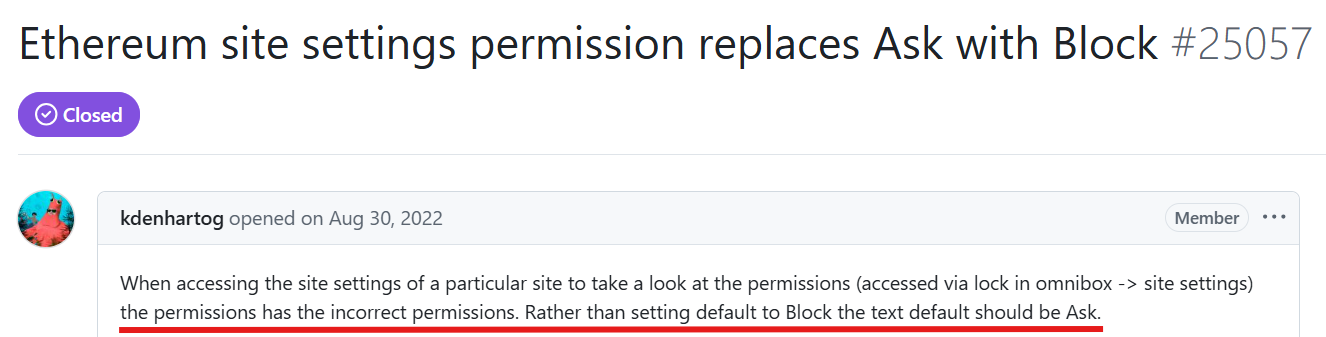}
        \includegraphics[width=\linewidth]{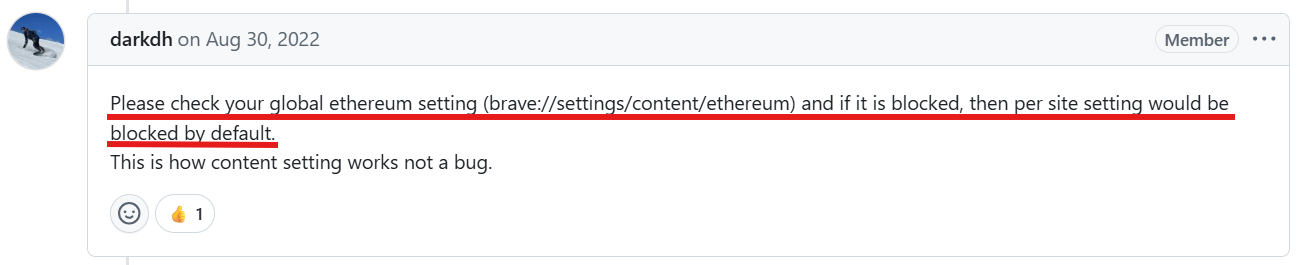}
        \caption[Faulty Configuration Issue]{Faulty Configuration Issue\footnotemark}
        \label{fig:faulty_config_example}
    \end{subfigure}
    
    \vspace{1em} 
    
    \begin{subfigure}{\linewidth}
        \centering
        \includegraphics[width=\linewidth]{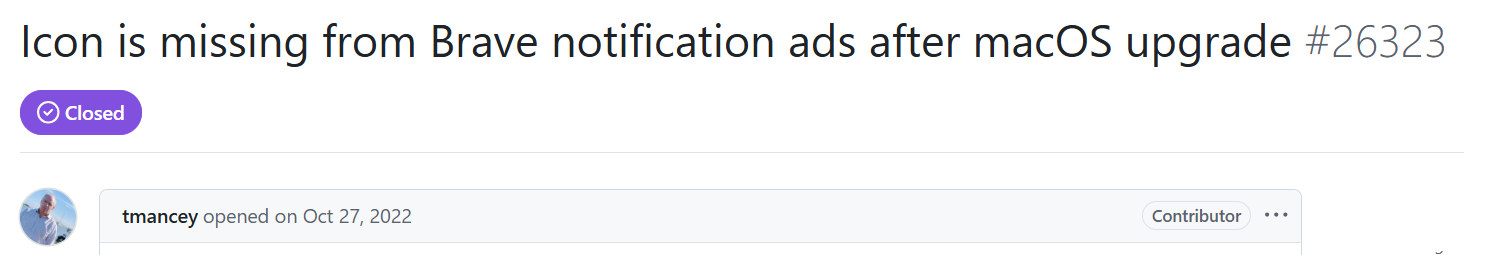}
        \includegraphics[width=\linewidth]{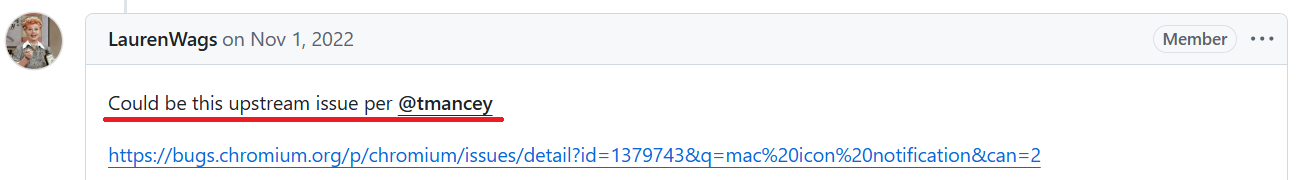}
        \caption[External System Dependency Issue]{External System Dependency Issue\footnotemark}
        \label{fig:feature_request_example}
    \end{subfigure}
    
    \caption{Invalid Bug Report Examples with Different Root Causes}
    \label{fig:placeholder}
\end{figure}
\addtocounter{footnote}{-1}
\footnotetext{\url{https://github.com/brave/brave-browser/issues/25057}}
\stepcounter{footnote}
\footnotetext{\url{https://github.com/brave/brave-browser/issues/26323}}

After identifying the root causes, the customer support team resolves these invalid bug reports using what we introduce as \textbf{\textit{no-code fixes}}, fixes that do not require changes to the source code. In this study, we argue that each potential root cause may require distinct no-code fixes. For instance, a user configuration error could be resolved by finding the required configuration on the user's machine (see Figure~\ref{fig:faulty_config_example}). In contrast, an external system dependency issue could be resolved by explaining that the issue originates from an upstream external issue (see Figure~\ref{fig:feature_request_example}). The customer support team can also consider these root causes while providing no-code fix solutions to the reporters.

Early detection and resolution of invalid bug reports could significantly reduce the workload of all stakeholders involved in the bug tracking lifecycle, including users, customer support, project managers/team leads, and developer teams. By identifying invalid bug reports early, the customer support team is relieved of the repetitive task of manually reviewing and classifying them. Instead, they can concentrate on verifying valid reports and reviewing automatically generated no-code fixes. Once the valid reports are classified, they are forwarded to the team leads, who then assign them to the appropriate developers. Except for a few edge cases, team leads usually prioritize valid issues that are more likely to require code changes \cite{yang2025enhanced}. Consequently, developers can devote their time to genuine software bugs without having to determine whether a bug report actually requires source code modification.

 However, this may still leave reporters uncertain about how to resolve the invalid bug reports. Incorporating no-code fixes directly addresses this gap by guiding reporters toward the appropriate solution. As a result, it could streamline the resolution process, increasing reporter satisfaction and engagement for future contributions \cite{torun2025past}.


Some earlier studies have applied different machine learning (ML) algorithms \cite{laiq2022_ttr, fan2020} to detect invalid bug reports. Some others used deep learning (DL) models \cite{he2020, li2022deeplabel} for the same task. More recent studies have applied large language models (LLMs) \cite{laiq2025}, while others have implemented hybrid solutions that combine DL and LLM approaches \cite{du2024llm} to detect invalid bug reports. In addition to basic validity detection, several studies have empirically investigated the root causes behind invalid bug reports \cite{laiq2023, su2017creating, panichella2021won, sun2011bug, herzig2013}. However, these works are limited because they rely on manual analysis rather than proposing automated methods for invalid subclassification by root causes. Although early studies exist in this field, they do not go beyond detecting invalid bug reports or merely discussing their root causes. Invalid bug reports need to be detected, categorized by potential root causes, and resolved with no-code fixes. To the best of our knowledge, no previous study has automatically detected the root causes of invalid bug reports and generated no-code fixes to resolve them. 

In our previous work \cite{dinc2025judge}, we employed a hybrid approach, using multiple ML models and feeding their results to a Judge LLM to determine the \textit{binary validity} of a bug report. The present study addresses a different task that begins where \cite{dinc2025judge} ends: given a report deemed invalid, it (i) subclassifies the report by its root cause and (ii) generates an actionable no-code fix. Its contributions are disjoint from \cite{dinc2025judge} in task (root-cause subclassification and resolution generation instead of binary validity detection), in data (a new manually annotated evaluation dataset), and in evaluation methodology (assessing generated no-code fixes with BERTScore and a Judge LLM against annotated ground truth). We do not include validity classification in this paper as it would shift the focus away from invalid bug report subclassification and no-code fix suggestion. The study also aims to evaluate and compare different approaches to achieve these objectives and validate their effectiveness.

To achieve these objectives, we address the following research questions (RQs):

\textbf{RQ1: How can invalid bug reports be subclassified according to their root causes?} \label{introRQ1}

With this research question, our aim is to identify the root causes of invalid bug reports automatically. 

\textbf{RQ1.1: How do LLMs perform in subclassification of invalid bug reports?}

With this research question, we aim to evaluate how different LLMs perform on the subclassification task of invalid bug reports based on their root causes.

\textbf{RQ1.2: Does context engineering affect the invalid bug report subclassification performance?}

With this research question, we aim to evaluate the effects of retrieval augmented generation (RAG) and agentic web search on the subclassification of the invalid bug reports with LLMs.

\textbf{RQ2: How can invalid bug reports be resolved with no-code fixes?}\label{introRQ2}

With this research question, our aim is to develop automated approaches to the crucial step of generating no-code fixes. 

\textbf{RQ2.1: How do LLMs perform in generating no-code fixes without subclass knowledge as prior information?}

With this research question, we aim to measure the performance of LLMs without the subclass information.

\textbf{RQ2.2: Does utilizing the root cause subclass information of invalid bug reports affect the no-code fix generation performance?}

With this research question, we aim to examine the impact of using the root cause subclassification of invalid bug reports (RQ1) on the generation of no-code fixes. 

\textbf{RQ2.3: Does context engineering affect the no-code fix generation performance?}

With this research question, we aim to evaluate the effects of RAG and web search tool utilization applications on no-code fix generation.


To address these RQs, we designed a multi-step evaluation framework called {\tool} that leverages LLMs to subclassify invalid bug reports and generate no-code fixes. We first established a baseline (vanilla) setting, evaluating the inherent capabilities of both proprietary and open-weight LLMs to map bug reports to predefined seven invalid subclasses: \textbf{External System \& Dependency Issues}, \textbf{Faulty Configuration}, \textbf{Feature Request}, \textbf{Non-reproducibility}, \textbf{Question}, \textbf{Working as Designed}, and \textbf{Wrong Version}. Then, {\tool} subsequently suggests actionable no-code fixes. To isolate the impact of subclassification on the quality of generated resolutions (RQ2.1 vs RQ2.2), we conducted an ablation study that requested no-code fixes entirely without subclass priors, later comparing these outcomes with generations informed by the root cause mappings.

Building upon the vanilla baseline, we implemented an RAG pipeline to investigate the role of external context in improving classification performance (RQ1.2) and no-code fix quality (RQ2.3). Our RAG system retrieves relevant developer wiki articles and similar historically invalid bug reports utilizing dense vector search. Furthermore, we utilize a recent agentic web search framework introduced by Li et al. \cite{li2025webthinkerempoweringlargereasoning} to test the impact of web search on LLM performance. To ensure robust evaluation across all experimental paradigms, the quality of the generated no-code fixes was automatically assessed using BERTScore \cite{zhang2019bertscore} and a Judge LLM to determine semantic and functional alignment with human-annotated ground truths. Although several classical text-matching metrics are established in the literature, they rely heavily on exact token overlaps and lack the capacity to evaluate deep contextual alignment. To avoid misleading or superficial performance variances, such metrics were intentionally excluded from our calculations.


The novelty of this study lies in the task formulation and the evaluation resource rather than in the underlying techniques: we introduce an automated system for subclassifying invalid bug reports into root-cause-oriented subclasses consolidated from prior empirical studies, and for further generating no-code fixes after subclassification, which was not available in the previous studies. To the best of our knowledge, there is no particular dataset that explicitly subclasses invalid bug report content by the root cause. We therefore also introduce a curated evaluation dataset comprising invalid bug reports, their invalidity subclasses, and the respective no-code fixes (if any). This dataset serves as ground truth for our study and possible future studies in this domain.

Our study has the following contributions:

\begin{enumerate}
    \item A root-cause-oriented invalid bug report subclass taxonomy, consolidated from prior empirical studies, together with an automated subclassification methodology.
    \item A novel approach to generate no-code fixes to resolve invalid bug reports.
    \item A comprehensive evaluation of the effect of prior information on the invalid bug report subclassification task.
    \item A comprehensive evaluation of different experimental settings on the no-code fix generation task. 
    \item A curated benchmark of invalid bug reports, their root-cause subclasses, and verified no-code fix suggestions.
\end{enumerate}


The remainder of this paper is organized as follows. \Cref{sec: relatedworks} reviews the relevant literature and discusses prior studies on invalid bug reports and their classification, as well as other points concerning no-code fix generation, highlighting the research gap. \Cref{sec: methodology} presents the curated benchmark and the methodology for the \textit{\tool} framework, detailing the design and workflow. \Cref{sec:results} reports the experimental results and the performance evaluation of different experimental settings. \Cref{sec: discussion} provides a detailed discussion of the findings and their implications. \Cref{sec: threats} presents the possible threats to validity. Finally, \Cref{sec: conclusion} concludes the paper by summarizing the key outcomes of the obtained results and suggesting directions for future work.

\section{Related Work and Background}
\label{sec: relatedworks}

In this section, we review the existing literature relevant to our study, establishing the foundational background on bug report validity and the root causes of invalid bug reports. We also examine the modern application of LLMs, with and without the additional context provided by RAG or web search. We inspect their usage in software maintenance and customer support workflows, highlighting the critical research gap in automated resolution for invalid bug reports.

\begin{table*}[htbp]
    \centering
    \caption{Root Causes of Invalid Bug Reports Mentioned by Source}
    \label{tab:actual_subclass_phrases}
    \renewcommand{\arraystretch}{1.5} 
    \begin{tabular}{@{} 
        >{\raggedright\arraybackslash}p{3.5cm} 
        >{\centering\arraybackslash}p{2.0cm}
        >{\centering\arraybackslash}p{2.0cm} 
        >{\centering\arraybackslash}p{2.0cm} 
        >{\centering\arraybackslash}p{2.0cm} 
        >{\centering\arraybackslash}p{1.5cm} 
        >{\centering\arraybackslash}p{1.5cm} 
    @{}}
        \toprule
        & \multicolumn{5}{c}{\textbf{Sources}} \\
        \cmidrule(l){2-6}
        \textbf{Identified Common Root Causes} & \textbf{Laiq et al. (2023) \cite{laiq2023}} & \textbf{Su et al. (2017) \cite{su2017creating}} & \textbf{Panichella et al. (2021) \cite{panichella2021won}} & \textbf{Sun (2011) \cite{sun2011bug}} & \textbf{Herzig et al. (2013) \cite{herzig2013}}\\
        \midrule
        External System \& Dependency Issues & \unchecked & Limitations and Behaviors of External Constraints & \unchecked & Error in External Systems & Issues Caused by Third Party Libraries \\
        Faulty Configuration \& Environment Setup & Faulty Configuration & Wrong Settings & Configuration Problem on the User Side & Wrong Configuration Parameters & \unchecked \\
        Feature Request & New Requirement & \unchecked & Feature Requests & Feature Requests & Feature Request \\
        Non-reproducibility & Non-reproducible & Irreproducible Defects & Not Replicable Bug & Reproducing Not Successful & \unchecked \\
        Question & \unchecked & \unchecked & Question & \unchecked & \unchecked \\
        Working as Designed & Working as Expected & Working as Designed & Wrong Usage of Functionality & Misunderstanding on Functionality & Does Not Comply with User Expectations \\
        Wrong Version (Already Fixed) & Wrong Version & \unchecked & Problem Already Fixed with the New Version & Build was Not New Enough to Contain the Fix & Versions are Open to Backport \\
        \bottomrule
    \end{tabular}
\end{table*}

\subsection{Bug Report Validity and Automated Triaging}\label{subsec: bugReportValidity}

The high volume of incoming bug reports necessitates robust automated triaging systems. Herzig et al. \cite{herzig2013} demonstrated that a significant percentage of bug reports are fundamentally misclassified by users, often acting as feature requests or routine tasks, which severely impacts the reliability of bug prediction models and wastes developer time. To mitigate this manual overhead, early empirical studies explored supervised \cite{supervised}, unsupervised \cite{unsupervised}, and semi-supervised \cite{semi_supervised} ML algorithms to automatically categorize bug reports. As the field advanced, researchers adopted DL to capture the semantic complexities of natural language descriptions \cite{qin2018classifying, ye2018bug, li2022deeplabel}.

However, treating bug validity purely as a binary classification stops short of addressing what aspects of the reports are missing for them to be valid. Recognizing this, further studies focused on extracting explanatory patterns, like the structural parts of the bug report or some keywords. He et al. \cite{he2020} proposed a DL-based approach to simultaneously determine validity and extract these patterns, while transformer architectures like BERT \cite{meng2023which} have been leveraged for similar rationale extraction.

These studies primarily treat validity as a routing or binary classification problem, stopping short of addressing the specific actions required once a report is deemed invalid.

Building on the need for actionable feedback, our previous system, Judge the Votes \cite{dinc2025judge}, extended beyond mere classification by providing users with automated, context-specific suggestions. Yet, this earlier work did not measure the quality of these fixes or subclassify invalid reports based on their root causes.

\subsection{Subclassification of Invalid Bug Reports}\label{subsec: subclassification}

Understanding \textit{why} a bug report is invalid is also crucial, as the root cause might dictate the subsequent resolution strategy \cite{laiq2023, fan2020}. A primary cause of invalidity is duplication, which has been extensively addressed by specialized word embedding models \cite{budhiraja2018towards} and DL techniques \cite{hybrid_deshmukh2017towards} designed to retrieve similar historical reports.

Beyond duplicates, reports are frequently marked invalid because they stem from distinct root causes, such as user misunderstandings, faulty environment configurations, or they are feature requests mistakenly reported as bugs. In order to define these root causes, researchers have conducted empirical analyses and proposed various taxonomies to subclassify invalid bug reports.

Sun \cite{sun2011bug} attributed many invalid bug reports to testing errors or a lack of system knowledge. Similarly, Su et al. \cite{su2017creating} found that 45\% of invalid defects are actually ``Working as Designed.'' Herzig et al. \cite{herzig2013} further highlighted this expectation gap, showing that over 33\% of reported bugs are simply misclassified feature requests or documentation issues. Panichella et al. \cite{panichella2021won} found that GitHub "wontfix" issues are often just support questions or out-of-scope requests. Laiq et al. \cite{laiq2023} used topic modeling (LDA) to extract broader patterns, identifying non-reproducibility, faulty environments, and user inexperience as key drivers of invalid bug reports. The overview of the mentioned invalid bug report root causes per source is displayed in \Cref{tab:actual_subclass_phrases}. None of the previous studies focused on fully automated invalid subclass detection. As bug tracking paradigms evolve into the Generative AI era \cite{torun2025past, akyol2025improbr}, subclassification has become a prerequisite for routing invalid bug reports away from human developers and toward intelligent support agents.

\subsection{LLMs and RAG in Software Maintenance}\label{subsec: rag4BugReports}

State-of-the-art LLMs have revolutionized automated bug triage by enabling sophisticated reasoning over technical documentation and bug reports \cite{wang2024large}. For instance, the \textit{Judge The Votes} framework \cite{dinc2025judge} evaluated the performance of LLMs in validity classification, leveraging model voting mechanisms and comparisons between similar issues to determine report legitimacy. Despite these capabilities, LLMs often hallucinate when lacking project-specific context, leading to inaccurate or generic resolutions \cite{kalai2025language, kiashemshaki2025automated}.

To address these limitations, recent studies have heavily adopted RAG to ground LLM outputs in repository realities \cite{tao2025retrieval}. By retrieving relevant documents before generation, RAG ensures that the model's responses are anchored in the specific codebase and history of a project. For example, \textit{RAG4Tickets} \cite{baqar2025rag4tickets} employs semantic embeddings to retrieve historical bug reports, passing them as context to an LLM to generate context-aware ticket resolutions.

RAG has also been applied to assess the quality and authenticity of the reports themselves. A 2026 study \cite{ren2025credibility} utilized RAG combined with zero-shot reasoning to successfully detect fabricated, AI-generated bug reports in bounty programs. Furthermore, tools like \textit{ImproBR} \cite{akyol2025improbr} use RAG pipelines alongside LLMs to automatically detect missing execution steps in bug reports and dynamically generate improved, reproducible instructions. While these advanced systems significantly reduce the maintenance burden by contextualizing and refining code-level bugs, they largely ignore the distinct workflow required for providing direct answers to invalid bug reports, such as configuration problems or conflicting user expectations.

\subsection{Agentic Web Search for Information Retrieval}\label{subsec:agentic-web-search}

Integrating web search agents addresses the siloed nature of traditional maintenance tools by treating the live web as dynamic external memory \cite{li2025webthinkerempoweringlargereasoning, zhang2025websearchagenticdeep}. Unlike RAG systems that rely on a static knowledge base and that can retrieve outdated or duplicate information, agentic web search, which is often framed as \textit{Agentic Deep Research} \cite{zhang2025websearchagenticdeep}, uses autonomous feedback loops to verify user-reported symptoms against real-time external sources such as upstream changelogs, third-party issue trackers, and community discussions. By going beyond internal wikis, these agents can uncover ``External Dependency'' or ``Wrong Version'' root causes that static systems would miss, providing the depth of context needed for large-scale software products.

Recent frameworks further show that technical triage often requires multi-step retrieval rather than a single query. Systems such as \textit{WebExplorer} \cite{liu2025webexplorerexploreevolvetraining} and \textit{BrowseMaster} \cite{pang2025browsemasterscalablewebbrowsing} scale tool usage and employ specialized agent architectures to support long-horizon reasoning on the web. In particular, \textit{WebExplorer} enables agents to iteratively refine queries and navigate complex web structures, interactively following links and filtering developer discussions to reconstruct the chain of events leading to the problem. Finally, \textit{WebThinker} \cite{li2025webthinkerempoweringlargereasoning}, the foundational framework used in our study, utilizes the ``Think-Search-and-Draft'' architecture, which allows an LLM to interleave its internal chain-of-thought with deep navigation of web elements.

\subsection{Bridging Customer Support and Bug Tracking with No-Code Fixes}\label{subsec: customerSupportBots}

As LLM capabilities expand, their role is shifting from passive classification tools to active resolution assistants \cite{ruan2025specrover, zhang2024autocoderover}. In a standard bug tracking framework, valid bug reports are routed to developers. However, invalid bugs, often stemming from user misunderstandings, faulty configurations, or external system dependencies, traditionally require time-consuming customer support intervention. This dynamic closely mirrors the challenges faced in consumer-facing applications, where a massive volume of user-reported issues are not actual system failures, but rather requests for usage workarounds or policy clarifications \cite{sudeep2024revolutionizing}. The fundamental nature of these consumer-facing queries is similar to the subclasses of invalid bug reports in software engineering; both scenarios involve users struggling with the intended design or configuration rather than encountering a source code defect \cite{fan2020}. Because consumer issues and invalid bug reports share these underlying root causes, they naturally necessitate a similar resolution process. Recent benchmarks like \textit{ECom-Bench} \cite{wang2025ecom} evaluate the capacity of multimodal LLM agents to resolve real-world customer support issues by simulating user interactions and providing immediate solutions. In app-centric ecosystems, instead of silently dismissing an invalid bug report, the AI agent should act as an advanced support bot \cite{torun2025past}, which actively resolves the issue by generating actionable \textit{no-code fixes}, such as step-by-step UI configuration guides, explanations of design choices, or workaround instructions. Applying this customer support resolution paradigm directly to software engineering allows organizations to automate the handling of invalid bug reports, which might improve first-contact resolution rates while keeping developers strictly focused on valid bug reports, which are source code defects.

\subsection{The Need for a No-Code Fix Benchmark}\label{subsec: nocodeFixBenchmark}

While extensive benchmarks exist for evaluating LLMs on source code generation and program repair, such as SWE-bench \cite{jimenez2023swe} and Livecodebench \cite{jain2024livecodebench}, there is a notable gap in benchmarking the generation of natural language no-code fixes. Recent large-scale datasets, such as \textit{GitBugs} \cite{patil2025gitbugs}, provide over 150,000 bug reports with resolution metadata, which greatly supports tasks such as duplicate detection and automated triaging. However, datasets like GitBugs do not explicitly label bug report content by invalid subclasses, nor do they assess the semantic correctness or user-helpfulness of the actual textual responses provided to reporters. The lack of a standardized benchmark for evaluating AI-generated workarounds limits the development of automated no-code fix systems. This gap motivates the creation of our novel benchmark to systematically evaluate how well AI can generate accurate, safe, and helpful no-code fixes for invalid bug reports.

\begin{table*}[t]
\caption{Taxonomy of Invalid Bug Reports: Root Causes, Descriptions, High-Level No-Code Fix Guidelines and Examples}
\label{tab:invalid_root_causes}
\centering
\small
\begin{tabularx}{\textwidth}{p{2cm}XXX}
\hline
\rowcolor[HTML]{EFEFEF}
\multicolumn{1}{>{\centering\arraybackslash}p{2cm}}{\textbf{Invalid Subclass (Root Cause)}} & \multicolumn{1}{c}{\textbf{Description}} & \multicolumn{1}{c}{\textbf{High-Level No-Code Fix Guideline}} & \multicolumn{1}{c}{\textbf{Example}} \\ \hline

External System \& Dependency Issues & 
Failure triggered by defects, outages, or limitations in third-party systems or hardware outside the maintainers' control. & 
Direct the user to the relevant third-party system and explain that the issue originates from an external dependency. &
\textbf{R: }\textit{Brave icon is missing from Notification Ads after macOS update.}

\textbf{M: }\textit{It is an \textbf{upstream issue} originating from Chromium.}~\cite{external_dependency} \\ \hline

Faulty Configuration \& Environment Setup & 
Incorrect user-side setting, parameter, or local environment mismatches before the execution. & 
Explain why the problem arises. Describe the required configuration and how to apply the necessary modifications. &
\textbf{R: }\textit{Tabs aren't visible or clickable after going full screen mode.}

\textbf{M: }\textit{Enabling \textbf{Always show toolbar in full screen} option will work.}~\cite{faulty_configuration} \\ \hline

Feature Request & 
User requests new functionality or enhancements rather than identifying a fault in the existing code. & 
Inform the user that the report is a feature request rather than a bug. If there is a workaround solution, explain it. Then, decide whether to implement the requested feature and notify users accordingly. &
\textbf{R: }\textit{Like Chrome \& Firefox, can Brave also have a development version?}

\textbf{M: }\textit{We have already Nightly, Beta and Dev versions. No need to proceed.}~\cite{feature_request} \\ \hline

Non-reproducibility & 
Cannot recreate the bug due to missing steps, race conditions, or one-time occurrences. & 
Ask the user for a detailed description, Steps to Reproduce (S2R), Expected Behavior (EB), Observed Behavior (OB), and complete environmental details if any of them are missing. If they are already known, but the bug exhibits intermittent behavior, suggest trying again at a different time. &
\textbf{R: }\textit{Crash happened in Brave Ads page.}

\textbf{M: }\textit{Blocked until we get \textbf{more information} regarding crashes.}~\cite{non_reproducibility} \\ \hline

Question & The report is an inquiry seeking help or discussion rather than a functional defect. & Provide a solution if available, or inform the user if the question cannot be answered. &
\textbf{R: }\textit{What format does Brave use to store date/time for Ads feature?}

\textbf{M: }\textit{Unix Epoch, thanks.}~\cite{question} \\ \hline

Working as Designed (Conflicting Expectation) & 
Software functions according to specifications, but the user perceives the correct behavior as a defect due to conflicting expectations. & 
Explain the system or specific feature in natural language. &
\textbf{R: }\textit{The browser keeps getting restarted after each update. I lose all my tabs.}

\textbf{M: }\textit{Changes only reflect when browser is restarted. This current behavior is \textbf{expected}.}~\cite{working_as_designed} \\ \hline

Wrong Version (Already Fixed) & 
Issue exists in an outdated or unsupported version but is already resolved in a newer release. & 
Inform the user that the issue has already been resolved in a newer version and kindly request that the user updates to that version. & \textbf{R: }\textit{While setting up Brave Wallet, the recovery phase shows duplicate options, indicating a possible bug.}

\textbf{M: }\textit{It was \textbf{already resolved} in version 1.35.x.}~\cite{wrong_version} \\ \hline

\end{tabularx}
\small{\textit{Note: The reporter (R) and the maintainer (M) actors have been referred in abbreviations at the \textit{Example} column.}}
\end{table*}

\begin{table*}[t]
\centering
\caption{Invalid Bug Report Labels, Descriptions, and Distribution in Brave Browser Repository \cite{brave_browser_labels}}
\label{tab:invalid-labels}

\small
\setlength{\tabcolsep}{6pt}
\renewcommand{\arraystretch}{1.2}

\begin{tabularx}{\textwidth}{cXc}
\hline
\textbf{Label} & \textbf{Description} & \textbf{Count} \\
\hline
closed/duplicate &
Issue has already been reported. &
2285 \\

closed/invalid & Not an issue with the browser. Possibly opened by mistake or there could be user error. &
2083 \\

closed/stale &
Issue is no longer relevant, often due to inactivity or deprecated functionality. &
1647 \\

closed/not-actionable & Closed because there is no way to reproduce the error or there is no clear action the team can take. &
985 \\

closed/wontfix & Working as intended. The reported behavior will not be fixed. &
925 \\

closed/works-for-me & Issue was closed because nobody was able to reproduce. &
712 \\

closed/no-milestone & No longer an issue. Not fixed by a particular release or milestone. &
297 \\

question & Seeking inquiries or clarifications. & 194
 \\

support & Addition/edits to articles on support website. &
66 \\

closed/workaround &
Issue has been resolved through a workaround and does not require a client-side fix. &
3 \\

closed/fixable-by-custom-rules &
Issue is fixable using custom rules or filter lists rather than code changes. &
1 \\

\hline
\textbf{Total} & & \textbf{8289} \\
\hline
\end{tabularx}
\end{table*}

\section{Methodology}
\label{sec: methodology}

In this section, we explain the curated taxonomy of invalid subclasses, the {\tool} Benchmark creation steps, and the {\tool} methodology.

\subsection{Invalid Bug Report Taxonomy Based on Root Cause} \label{subsec: invalidTaxonomy}

 The literature review leads us to a consolidated taxonomy based on root causes for invalid bug reports. Unlike generic labels mostly used in GitHub repositories, {\tool} Benchmark prioritizes the \textbf{root causes} of invalidity. The final clusters of invalid subclasses are \textbf{External System \& Dependency Issues}, \textbf{Faulty Configuration  \& Environment Setup}, \textbf{Feature Request}, \textbf{Non-reproducibility}, \textbf{Question}, \textbf{Working as Designed (Conflicting Expectation)},  and \textbf{Wrong Version (Already Fixed)}. While previous studies \cite{budhiraja2018towards, hybrid_deshmukh2017towards} often classified Duplicate reports as a subclass of invalid bugs, we exclude them from our taxonomy. This decision is twofold: first, a duplicate may still represent a valid bug that requires actual code modifications. Second, duplicates are typically resolved simply by linking to a parent issue rather than offering a distinct solution. Because our study focuses on extracting unique, context-aware no-code fixes, duplicate reports fall outside the scope of our analysis. The mapping definitions of each invalid subclass per source are given in Table~\ref{tab:actual_subclass_phrases}. As detailed in Table~\ref{tab:invalid_root_causes}, the taxonomy organizes these reports by three primary dimensions:

\begin{itemize}
    \item \textbf{\textit{Invalid Subclass (Root Cause)}}:  Represents the root cause of the invalidity.
    \item \textbf{\textit{Description}}: Outlining the specific symptoms or traits of the report.
    \item \textbf{\textit{High-Level No-Code Fix Guideline}}: Identifies actionable, non-invasive solutions such as configuration guidance or workarounds.
\end{itemize}

By applying this taxonomy to the real-world bug reports, we can effectively distinguish between the root causes, such as configuration mismatches, conflicting user expectations, and external dependencies. This structured approach ensures that the final benchmark is not merely a collection of invalid bug reports, but a categorized dataset optimized for automated resolution by invalid subclass.

\subsection{{\tool} Benchmark}
\label{sec: dataset}

\subsubsection{Data Selection \& Fetching}

To achieve our primary objective of subclassifying invalid bug reports and generating no-code fixes, we first established a method to distinguish valid reports from invalid ones. Our dataset focuses on user-facing open-source repositories on GitHub, which could provide a rich source of bug reports.

Before detailing our dataset, it is important to clarify the terminology used in this ecosystem. In the context of GitHub, \textbf{GitHub Issues} correspond directly to \textbf{bug reports}. Throughout this paper, we treat GitHub Issues as the primary unit of analysis for bug reports, and we use the two terms interchangeably.

To ensure high data quality and relevance, we established strict static criteria for selecting our target repository:

\begin{figure*}[t]
    \centering
    \includegraphics[width=1.0\textwidth]{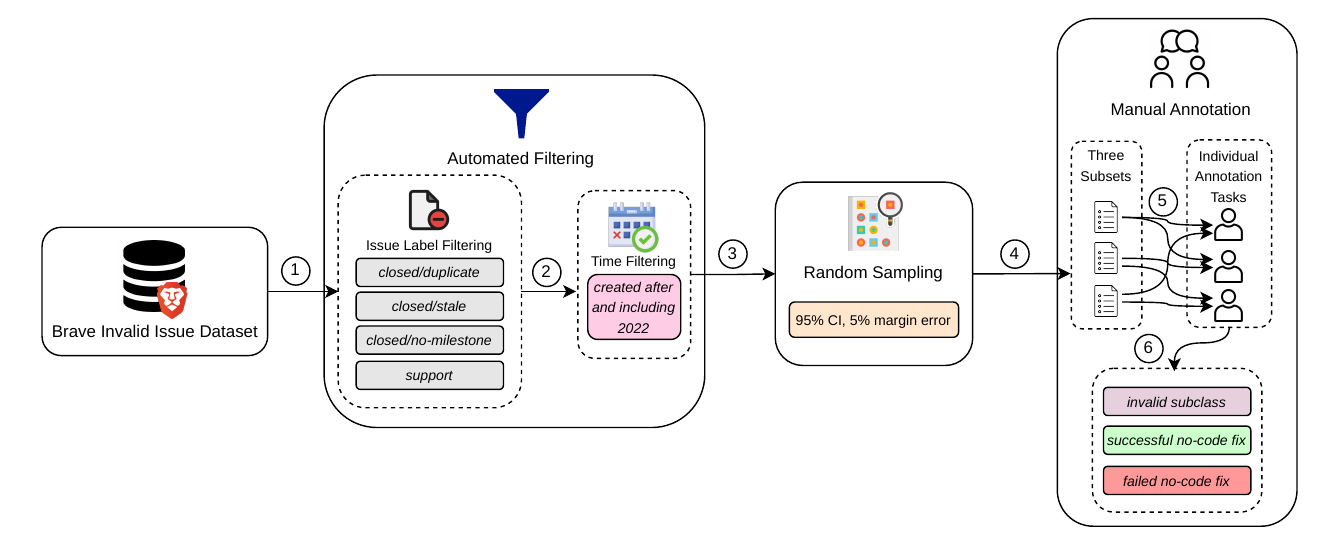}
    \caption{Overview of Evaluation Benchmark Curation Workflow}
    \label{fig:datasetCuration}
\end{figure*}

\begin{itemize}
    \item \textbf{Popularity:} The repository must have at least 20,000+ stars, ensuring it is a widely used and active project.
    \item \textbf{Volume:} The repository must contain 40,000+ bug reports to provide a statistically significant dataset.
    \item \textbf{Labeling:} Crucially, the repository must actively use a set of labels that explicitly categorize \textbf{invalid} bug reports.
\end{itemize}

After conducting research based on these parameters, we identified the \textbf{Brave}\footnote{\url{https://github.com/brave/brave-browser}} browser repository as the ideal candidate. It not only met our volume and popularity thresholds but also maintained a rigorous labeling system for subclassifying invalid issues, making it highly suitable for our classification goals.

By mining the Brave repository with the GitHub API, we constructed a comprehensive dataset comprising \textbf{38,789} closed bug reports (captured as of January 19, 2026). Following the criteria specified in the Brave Browser wiki\footnote{{\url{https://github.com/brave/brave-browser/wiki/Issue-with-missing-milestones}}}, we identified \textbf{\invalidTotalCount} reports as \textbf{invalid} based on labels that the project considers \textit{invalid}. The distribution of labels within the invalid subset, with their explanations and counts, is detailed in Table~\ref{tab:invalid-labels}. Please note that because a single bug report can have multiple labels, the total number of labels does not match the number of invalid bug reports.

\subsubsection{Evaluation Benchmark}

The Evaluation Benchmark is a curated subset of invalid bug reports designed to provide a high-quality foundation for our analysis. While the initial extraction from the Brave repository yielded \invalidTotalCount\ reports, their native labeling system often reflects implementation-specific details rather than the fundamental reasons for invalidity. To bridge this gap, we implemented a curation pipeline that re-evaluates these reports based on the previously defined root-cause oriented \textbf{invalid subclass taxonomy}. The comprehensive workflow for this benchmark curation and manual labeling methodology is illustrated in Figure~\ref{fig:datasetCuration}.

\paragraph{Automated Filtering}
 Starting with \invalidTotalCount~ invalid bug reports, we have decided to filter out some of the Brave-native invalid labels mentioned in Table~\ref{tab:invalid-labels}, as some of those labels' objectives are not aligned with this paper's goals. Here are the reasons for the filtered-out labels.

\paragraph*{\textbf{closed/duplicate}} Duplicate reports were removed as they typically point to existing tickets rather than requiring unique no-code fixes.

\paragraph*{\textbf{closed/stale}}
Excluded due to prolonged inactivity and incomplete comment threads. While these issues became irrelevant over time, this does not imply invalidity; they simply lack sufficient information to definitively determine their actual validity.

\paragraph*{\textbf{closed/no-milestone}}
Excluded because they represent valid issues that were resolved in subsequent releases without being assigned to a specific milestone. Thus, they cannot be attributed to invalidity or user-side errors.

\paragraph*{\textbf{support}}
Excluded because maintainers typically redirect these reporters to support channels rather than discussing issue resolution. Consequently, these reports lack the technical context and root cause analysis required to evaluate their validity.
\\\\
After these exclusions, the filtered dataset consists of \textbf{\filteredInvalidCount} invalid bug reports.\\

Furthermore, among these \filteredInvalidCount~invalid bug reports, we decided to keep the reports created after and including \thresholdYear~i.e., issues opened at least six years after the repository's first bug report in 2016.. The main reasons for applying this filter are:

\begin{itemize}
    \item \textbf{Rich Knowledge Base and Discussion History}: Over several years, the Brave community has generated a vast corpus of technical discussions and community-driven solutions. By focusing on more recent reports, we leverage this established knowledge base, which provides the depth of context necessary to identify and reconstruct recurring no-code fixes.
    \item \textbf{Project Maturity and Maintenance Phase}: As the project reached a mature maintenance phase, it adopted standardized reporting templates and established a stable code base.
\end{itemize}

The filtered dataset consists of \textbf{\invalidReportCount}~invalid bug reports. 

\paragraph{Random Sampling}


To construct our evaluation benchmark, we conducted a manual annotation process on a subset of the data. Using Cochran's sample size formula \cite{cochran2007sampling}, we randomly selected \textbf{\invalidSampleCount} reports from the total pool of invalid bug reports ($N=\invalidReportCount$), a sample size that permits estimating population proportions over this pool with a 95\% confidence level and a 5\% margin of error. We note that this sampling guarantee applies only to such proportion estimates (e.g., the label-noise rates reported in Table~\ref{tab:final_label_annotation}); it does not confer statistical significance on comparisons between classification pipelines, and since the in-taxonomy subset after outlier exclusion contains 188 reports, the stated precision does not extend to per-subclass estimates. This sampled subset allows for an assessment of both invalidity subclassification and the extraction of proposed no-code fixes.

\paragraph{Manual Annotation Task Assignment}

The manual validation of the \invalidSampleCount~sampled invalid bug reports was conducted by three authors using a rotating-pair configuration. The dataset was partitioned into three subsets, with each subset assigned to a different pair of authors for independent review. This distribution resulted in each author assessing two-thirds of the total sample. In cases where the two primary annotators assigned conflicting labels, the third author who had not participated in the initial assessment of that specific subset reviewed the report to make the final determination.

\paragraph{Manual Annotation}
Before finalizing the rules and steps of the manual evaluation, the three authors initially manually annotated seven random invalid bug reports independently. Then, the labeling reasoning and perspectives have been discussed to finalize the manual annotation keypoints for a robust annotation process.

During the manual analysis, we considered several key data points in the analysis process:
\begin{itemize}
    \item \textbf{Issue Metadata:} We examined the title and description for clarity and quality of the demonstration of the issue, alongside the original labels assigned by developers and temporal events (e.g., the duration between comments).
    \item \textbf{Content and Artifacts:} We evaluated the comment threads for technical quality and relevance, including attachments like screenshots, screen recordings, and GitHub-specific reactions (e.g., \textit{thumbs up}, \textit{rocket} emojis) which signal community consensus.
\end{itemize}

\paragraph{Annotation Granularity}
The manual annotation was conducted at two levels:
\begin{enumerate}
    \item \textbf{Report-Level Classification:}
        \begin{itemize}
        \item \textit{Subclassification of Invalids:} Each report is assigned a single invalid subclass according to the taxonomy defined in Table~\ref{tab:invalid_root_causes}, identifying the specific root cause of its invalidity.
        \item \textit{Outlier Annotation:} To ensure the integrity of the benchmark and account for real-world edge cases, we introduced three outlier labels:
        \begin{enumerate}
        \item \textit{Valid}: Assigned to reports that, upon manual review, are found to represent actual software defects requiring a code change.
        \item \textit{No Conclusion}: Reserved for reports where the provided content and subsequent comment threads do not contain sufficient information to reach a definitive classification.
        \item \textit{Developer Workflow}: Reserved for reports that are opened for internal maintenance tasks that only include the details, such as the task description and acceptance criteria. 
        \item \textit{Others}: Reserved for reports that fall outside the scope of the established classification taxonomy.
        \end{enumerate}

        While our curation pipeline aims to isolate invalid bug reports, including these categories ensures the benchmark remains a transparent and representative reflection of the repository's reporting environment.
        \end{itemize}
        
    \item \textbf{Comment-Level Extraction:}
        \begin{itemize}
        \item \textit{No-Code Fix Identification:} We isolated individual comments that proposed no-code fix suggestions to the reported issue. Unlike the report-level classification, which is mutually exclusive, this is a multi-label task; a single bug report may contain multiple distinct comments identified as potential no-code fixes.
        \item \textit{Solution Verifiability:} To evaluate the reliability of these interventions, we extracted both \textbf{successful} and \textbf{failed} suggestions based on the following criteria:
        \begin{itemize}
        \item \textbf{Successful Fixes:} Comments where the proposed workaround or clarification was explicitly validated. Evidence of success includes: (1) direct acknowledgment from the reporter via text or reaction (e.g., a thumbs-up emoji), (2) the issue is immediately closed following the suggestion and not opposed by the reporter, or (3) a maintainer closing the issue while citing the specific comment as the resolution, such as \textit{'Issue closed based on this workaround...'}.
        \item \textbf{Failed Fixes:} Comments containing suggestions that were attempted but ultimately rejected. These are identified by subsequent feedback from the reporter or maintainers indicating the solution was ineffective, such as \textit{'This did not work'}.
        \end{itemize}
        \end{itemize}
\end{enumerate}

\paragraph{Inter-rater Reliability}
We evaluated the agreement between the annotators using Cohen’s kappa ($\kappa$) for invalid subclass annotation and the Jaccard similarity coefficient for the unstructured no-code fix comment extraction. The resulting scores were $\kappa = \textbf{0.5732}$, for all assessed bug reports,  with a moderate agreement \cite{landis1977} for \textit{invalid subclass} annotation. For \textit{no-code fix comment extraction}, we observed a Jaccard similarity of $J_s = \textbf{0.7322}$ for successful no-code fixes and $J_f = \textbf{0.9338}$ for failed no-code fixes, confirming high agreement on the selection of the no-code fix content. We note that moderate chance-corrected agreement on the subclass labels indicates that a portion of root-cause assignments are genuinely ambiguous even for human annotators. Although all disagreements were resolved through third-annotator judge, this agreement level places a practical ceiling on the model performance measurable against the resulting labels, since some model ``errors'' fall within the range of human disagreement. Accordingly, we refer to the adjudicated labels as a curated ground truth rather than an error-free gold standard.

\begin{table}[htbp]
    \centering
    \small 
    \renewcommand{\arraystretch}{1.2} 
    \caption{Manual Annotation Results for the Sampled Invalid Bug Reports}
    \label{tab:final_label_annotation}
    \begin{tabularx}{\columnwidth}{>{\raggedright\arraybackslash}X cccc} 
        \toprule
        \textbf{Subclass} & \textbf{Count} & \thead{\textbf{Percentage}} & \thead{\textbf{Having} \\ \textbf{Successful} \\ \textbf{No-Code Fix}} & \thead{\textbf{Having} \\ \textbf{Failed} \\ \textbf{No-Code Fix}} \\
        \midrule
        Working as Designed & 48 & 15.89\% & 48 & 0 \\
        Feature Request & 41 & 13.58\% & 26 & 0 \\
        Non-reproducibility & 35 & 11.59\% & 21 & 0 \\
        External System \& Dependency Issues & 20 & 6.62\% & 20 & 3 \\
        Faulty Configuration \& Environment Setup & 19 & 6.29\% & 19 & 2 \\
        Question & 13 & 4.30\% & 13 & 2 \\
        Wrong Version & 12 & 3.97\% & 12 & 0 \\
        \midrule
        \textbf{Base Annotations Total} & \textbf{188} & \textbf{62.25\%} & \textbf{159} & \textbf{7} \\
        \midrule
        Valid & 60 & 19.87\% & -- & -- \\
        Developer Workflow & 29 & 9.60\% & -- & -- \\
        No Conclusion & 23 & 7.62\% & -- & -- \\
        Others & 2 & 0.66\% & -- & -- \\
        \midrule
        \textbf{Outlier Annotations Total} & \textbf{114} & \textbf{37.75\%} & -- & -- \\
        \midrule
        \textbf{Total} & \textbf{302} & \textbf{100.00\%} & \textbf{159} & \textbf{7} \\
        \bottomrule
    \end{tabularx}
\end{table}

\paragraph{Final Benchmark \& Observations}

Following this process, we established a manually adjudicated evaluation benchmark containing both verified invalid subclasses and no-code fixes. The final invalid subclass distribution is given in Table~\ref{tab:final_label_annotation}. Based on the labeling results, 114 \textbf{(37.75\%)} of the sampled bug reports do not belong to the pre-defined invalid subclass taxonomy, where 60 \textbf{(19.87\%)} of these bug reports are labeled as valid, 23 \textbf{(7.62\%)} of them are labeled as no conclusion and two (\textbf{0.66\%}) of them were labeled as duplicate, which are considered as \textbf{Others}. This distribution highlights the practical inconsistency between the invalid label definitions and the final label annotation.

Furthermore, the benchmark details the distribution of no-code fixes across the invalid subclasses. Out of the 188 base annotations, a total of 159 successful and seven failed no-code fixes were identified. Notably, there is a pronounced need for no-code fixes in specific subcategories. Excluding Non-reproducibility and Feature Request, every single report in the remaining five invalid subclasses contained a successful no-code fix, reinforcing that their definitions heavily depend on how the fix resolves the issue. Interestingly, while External System, Faulty Configuration, and Question represent mid-to-lower overall volumes, they accounted for all seven of the failed no-code fix attempts in the dataset. This phenomenon underscores the potential complexity of providing actionable no-code fix for these specific operational, environmental, and design-related issues.

Particular patterns and outliers across the bug reports have emerged after manual annotation and constructive discussions between the authors. 

\noindent\rule{\linewidth}{0.4pt}

As an example of the outlier pattern, the manually annotated two \textit{duplicate} bug reports do not have the original Brave issue label \textit{closed/duplicate}.
\smallskip

{\footnotesize \textbf{Example:} In issue \#24089\footnote{https://github.com/brave/brave-browser/issues/24089}, the contributor 'srirambv' directly asks, '\textit{Dupe of \#23898?}' and the author confirms, '\textit{yep, that is a dupe},' even though the only \textbf{invalid Brave label} on the bug report was '\textit{closed/invalid}'.}

\noindent\rule{\linewidth}{0.4pt}

One of the sub-patterns observed inside the \textit{no conclusion} subset is that the issue remains stale for a subsequent time and then the particular feature related to the bug report gets deprecated in later updates. 

\smallskip

{\footnotesize \textbf{Example:} In issue \#21011\footnote{https://github.com/brave/brave-browser/issues/21011}, the reporter claims that the IPFS feature inside the Brave Browser has flyout menus and by quote '\textit{flyout menu items should have an ellipsis, as they require further user input}', which was reported in February 10, 2022. However, the issue remains stale until September 2, 2024 without any kind of clarification or suggestion about the issue. In September 2, 2024; the collaborator '\textit{vadimstruts}' responds by quote '\textit{The IPFS local node and scheme has been deprecated}.' Since the particular bug report remained stale for nearly 2.5 years without any fix suggestions or clarifications; no one can possibly understand whether the issue is valid or invalid.}

\noindent\rule{\linewidth}{0.4pt}

A common pattern among Brave maintainers is the creation of ``follow-up'' issues that lack self-contained context. 

\smallskip

{\footnotesize \textbf{Example:} Issue \#25842\footnote{https://github.com/brave/brave-browser/issues/25842} addressed an 81\% failure rate in crash dump uploads on the Brave Stable channel caused by an incorrect sampling probability. Although a patch was deployed and the issue is closed on October 20, 2022, issue \#26148\footnote{https://github.com/brave/brave-browser/issues/26148} was opened that same day as a direct follow-up. That issue particularly identified that the original problem persisted during frequent, consecutive crashes. However, the child issue relied entirely on the parent issue for context, by quote, '\textit{This is follow-up for \#25842.}'.}

\smallskip

The poor maintenance choice is that the child issue is inherently coupled with the parent issue, and the good practice should have been the parent issue being \textbf{re-opened} after another edge-case occurs \cite{kucuk2023characterizing}.

\noindent\rule{\linewidth}{0.4pt}

The other pattern is about the ground-truth no-code fixes. Some of the no-code fix comments left by maintainers tend to reference another bug report without sufficient explanation on what the final conclusion is.

{\footnotesize \textbf{Example:} Issue \#33511\footnote{https://github.com/brave/brave-browser/issues/33511} is about the ``Select and Search'' function of the ``Lazy Chrome'' extension, which stopped working after they updated to Brave version 1.59. However, the same issue has been opened in the Lazy Chrome repository\footnote{\url{https://github.com/frog1014/lazy\_chrome/issues/22}} and the main solution discussions took part in that issue report. As a result, in the Brave issue report, the author responds, by quote, '\textit{Figured out due to change in brave's extension permission control frog1014/lazy\_chrome\#22}'.}

\noindent\rule{\linewidth}{0.4pt}

The curated benchmark serves as a valuable asset for the software engineering research community. By providing a ground truth set of invalid bug reports paired with their specific no-code solutions, it enables the rigorous evaluation of automated classification models and solution recommendation systems, facilitating the development of tools that can alleviate the maintenance burden in open-source projects.

\subsection{{\tool} Methodology}

\begin{figure*}
    \centering
    \includegraphics[width=0.9\textwidth]{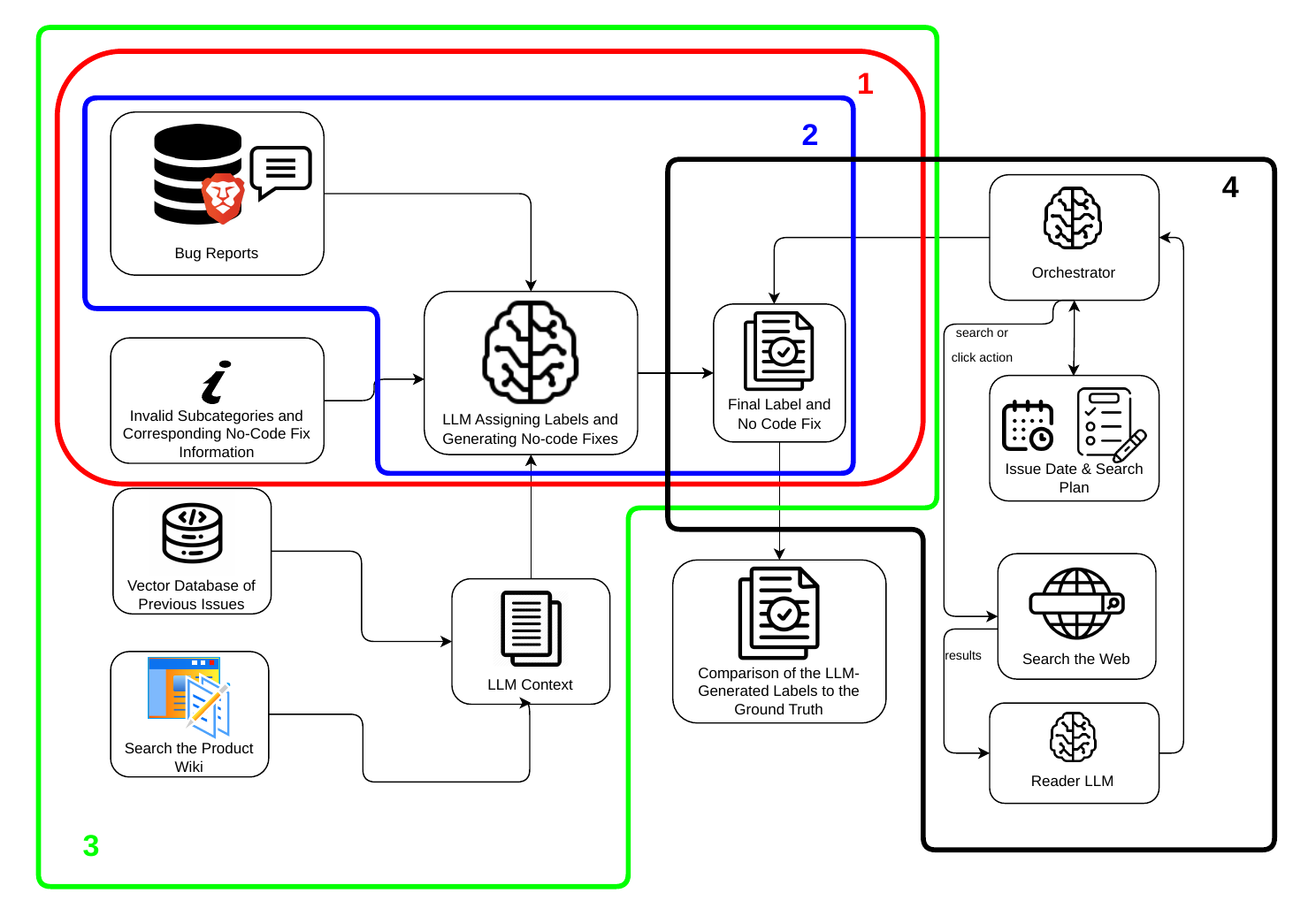}
    \caption{Overview of {\tool} Methodology}
    \label{fig:methodology}
\end{figure*}

We experimented with four distinct methodologies: (1) \textit{Vanilla LLM Pipeline}, (2) \textit{Vanilla LLM Pipeline Without Prior Invalid Subclass Information}, (3) \textit{RAG Pipeline}, and (4) \textit{Agentic Web Search Pipeline}. The tested methodologies return one invalid subclass and one suggested no-code fix, except for (2), and differ based on the tools and sources they use. Figure \ref{fig:methodology} presents the overall methodology.

\smallskip

\subsubsection{Prompt Configurations}

All mentioned approaches utilize \textbf{zero-shot} prompting strategy, consisting of two distinct components: a system prompt and a user prompt. 

The system prompt constructs the task's context by formally defining both an \textit{invalid bug report} and a \textit{no-code fix}. It instructs the model to classify the given report into a specific invalid subclass and generate an appropriate no-code fix. To guide this reasoning, the prompt incorporates the subclass descriptions and high-level no-code fix guidelines outlined in Table~\ref{tab:invalid_root_causes}, except for approach (2). Finally, the model is strictly instructed to return its response in a structured JSON format with the given three fields: \textit{classification} (except for (2)), \textit{reasoning}, and \textit{no\_code\_fix}. Since approach (2) is for ablation study purposes, it lacks subclass descriptions and guidelines for the no-code fixes in the system prompt.

Conversely, the user prompt serves strictly as the data input layer, injecting only the raw markdowns of \textit{title} and \textit{body} of the bug report for vanilla LLM pipelines, and additionally \textit{retrieved information} for RAG and agentic web search pipelines. The complete base system and user prompt templates are available in the \texttt{prompt.py} file of our replication package\footnote{\url{https://figshare.com/s/dc79aaf924dac61a1095?file=64399383}}.

\subsubsection{Vanilla LLM Pipeline}
The \textit{Vanilla LLM} pipeline serves as our primary zero-shot baseline.  In this setup, we have evaluated three proprietary and three open-source LLMs.

These LLMs were selected based on their \textbf{Intelligence Index} from Artificial Analysis\footnote{https://artificialanalysis.ai/methodology/intelligence-benchmarking}. This index evaluates core capabilities, such as \textit{reasoning} and \textit{knowledge}, by aggregating results from diverse datasets. According to the leaderboard from March 25, 2026, selected proprietary models are Gemini 3.1 Pro Preview, GPT-5.4 (xhigh), and Claude Opus 4.6 (max), and the open-source models are GLM-5 \cite{zeng2026glm}, MiniMax M2.7 \cite{minimax2026m27}, and Kimi 2.5 \cite{team2026kimi}. In this setting, we picked the best-performing open-source and proprietary LLMs based on their overall \textbf{vanilla invalid subclassification performances} for use in other experimental setups, , since subclassification constitutes the first stage of the pipeline and gates the downstream fix generation; we discuss the implications of this selection criterion in \Cref{sec: threats}.

\smallskip

\subsubsection{Without Prior Invalid Subclass Information}
To rigorously quantify the significance of our proposed taxonomy and structured guidelines, we formulated the \textit{Without Prior Information} ablation pipeline. This setup mirrors the Vanilla LLM pipeline; however, all invalid subclass priors are deliberately removed from the system prompt. The LLM is neither provided with the definitions of the seven invalid subclasses nor the tailored guidelines for generating no-code fixes. Instead, the model is prompted generically as an ``expert software engineer'' and must deduce a no-code fix resolution relying solely on its intrinsic, pre-trained knowledge.

\smallskip

\subsubsection{RAG Pipeline}
The \textit{RAG} pipeline is designed to ground the LLM's deductive reasoning in historical project data and project wiki. We populated two distinct vector databases: one containing \textbf{historical bug reports} and another containing \textbf{official Brave Wiki documentation}\footnote{\url{https://github.com/brave/brave-browser/wiki}}. We utilized the \texttt{zembed-1}\footnote{\url{https://zeroentropy.dev/articles/introducing-zembed-1-the-worlds-best-multilingual-text-embedding-model/}} text embedding model, selected from the Agentset embedding models leaderboard\footnote{\url{https://agentset.ai/embeddings}}, to generate vector embeddings. These embeddings are stored in \texttt{LanceDB}\footnote{\url{https://www.lancedb.com/}}, chosen for its efficient metadata filtering, which is critical for our chronological data leakage prevention. 

When evaluating a new bug report, the system uses the report's title and body as the query for a dense vector search. We employ a two-stage retrieval process: first, the system retrieves the top $k=20$ most semantically similar candidates from each database. To strictly prevent data leakage, historical bug reports are chronologically filtered using a metadata constraint (\texttt{created\_at < current\_created\_at}) to ensure only prior issues are retrievable. 

These initial candidates are then processed by the \texttt{zerank-2}\footnote{\url{https://zeroentropy.dev/articles/zerank-2-advanced-instruction-following-multilingual-reranker/}} cross-encoder re-ranking model, also selected from the Agentset leaderboard\footnote{\url{https://agentset.ai/rerankers}}. The reranker refines the selection to the top $k=5$ most relevant contexts. These highest-ranked contexts are appended to the LLM prompt, providing the contextual data the model needs to formulate its final classification and no-code fix.

\smallskip

\subsubsection{The Agentic Web Search Pipeline}\label{subsec:agentic-search-methodology}

To incorporate dynamic external evidence, we implemented an agentic web search pipeline loosely inspired by WebThinker-style reason--search--observe loops \cite{li2025webthinkerempoweringlargereasoning}. Given a bug report title, body, and creation date, the pipeline first constructs a compact research brief containing the issue context and the information need. An orchestrator LLM then performs a bounded information-gathering loop: it proposes natural-language search queries, retrieves web results through the Serper Search API \cite{serper2026}, removes the canonical original issue page to mitigate direct leakage, and summarizes the retained evidence with source URLs. When useful, the agent can also request a page visit; fetched page content is compressed into the same evidence format.

The search trajectory is constrained to limit cost, latency, and noise. In our experiments, the outer loop was capped at 35 turns and at most three search actions; page visits were treated as part of the same bounded trajectory. Search results were obtained from the live web index at experiment time, rather than reconstructed from the issue creation date, which we discuss as a threat to validity. Tool observations were concatenated in chronological order and passed as retrieved context to the downstream classifier/generator, which then produced the final invalid subclass and no-code fix.

All language-model calls in the browsing orchestrator were issued through a single OpenAI-compatible client configured to OpenRouter \cite{openrouter_api_2026}. For the OpenRouter-Kimi configuration, orchestration used \texttt{moonshotai/kimi-k2.5} with the default sampling temperature and a large token budget, while the final classifier used the same backbone with a lower temperature, the same token budget, and JSON-object response formatting when supported by the endpoint. Gemini-only agentic runs used \texttt{google/gemini-3.1-pro-preview}, with \texttt{google/gemini-2.5-flash} for query generation, page compression, and evidence summarization.

\subsection{Evaluation}

To evaluate the effectiveness of the proposed methodologies, we executed each pipeline in three independent trials to reduce the variance that can be caused by the probabilistic nature of LLMs \cite{niimi2025simple_3runs}. We kept the backbone LLMs at their default temperature settings to take advantage of their inherent reasoning capabilities, avoiding the constraints of the zero-temperature configuration. 

The evaluation is structured to address our two primary research questions:

\smallskip

\textbf{In RQ1,} we assess the performance of the invalid subclassification task using the \textbf{weighted F1-Score}, which provides a reliable measure given the imbalanced distribution of invalid subclasses. For each bug report, the final classification label is determined by majority vote across the three independent runs.

\textbf{In RQ2,} in order to evaluate the semantic and functional utility of the generated no-code fixes, we employ a dual-metric approach. 

\begin{itemize}
    \item \textbf{BERTScore F1:} This metric provides a token-level semantic similarity analysis. We compute this metric by comparing model outputs specifically against the successful ground-truth fix, since semantic closeness to the failure case is not the target objective.
    \item \textbf{Judge-LLM Evaluation:} To assess practical alignment, we utilize an LLM-as-a-judge framework. The judge role is fulfilled by \texttt{google/gemini-3.1-pro-preview}, selected for its strong instruction-following and reasoning capabilities, operating at its default temperature setting. The Judge LLM is presented with a \textit{contrastive three-part prompt} consisting of the predicted fix, a successful ground-truth no-code fix, and a failed ground-truth no-code fix. This allows the judge to perform a qualitative comparative assessment, determining whether the model's suggestion aligns with successful resolution patterns or mimics known failure modes.
\end{itemize}

For the evaluation in RQ2, we constrain our analysis to a subset of 159 bug reports that possess successful ground-truth no-code fixes, as ground-truth data is unavailable for the remainder of the benchmark.

\section{Results}
\label{sec:results}

We structurally organize our experimental findings to address our primary research questions regarding the invalid bug report subclassification and the no-code fix generation approach. To facilitate the reproducibility of our findings and support future research in automated subclassification and no-code fix generation for invalid bug reports, our complete replication package is publicly available at \url{https://figshare.com/s/dc79aaf924dac61a1095}.

\begin{table*}[htbp]
\centering
\caption{Comprehensive Subclassification Performance: Vanilla, RAG, and Agentic Web Search (Weighted F1)}
\label{tab:merged_classification_performance}
\resizebox{\textwidth}{!}{%
\begin{tabular}{@{}lcccccccccc@{}}
\toprule
\textbf{Subclass / Pipeline} & \textbf{Gemini 3.1} & \textbf{GPT-5.4} & \textbf{Opus 4.6} & \textbf{Kimi K2.5} & \textbf{GLM-5} & \textbf{Minimax M2.7} & \textbf{RAG+Gemini} & \textbf{RAG+Kimi} & \textbf{Agentic+Gemini} & \textbf{Agentic+Kimi} \\ \midrule
External System \& Dep.   & 0.62 & 0.56 & \textbf{0.71} & 0.47 & 0.49 & 0.46 & 0.67 & 0.42 & 0.57 & 0.43 \\
Faulty Config \& Env.     & \textbf{0.34} & 0.30 & 0.23 & \textbf{0.34} & 0.24 & 0.07 & 0.27 & 0.26 & 0.32 & 0.29 \\
Feature Request           & \textbf{0.81} & 0.76 & 0.74 & 0.77 & 0.75 & 0.75 & 0.79 & 0.76 & 0.77 & 0.79 \\
Non-reproducibility       & 0.79 & 0.81 & 0.77 & 0.63 & 0.78 & 0.66 & \textbf{0.85} & 0.76 & 0.76 & 0.72 \\
Question                  & 0.76 & 0.75 & 0.71 & 0.73 & 0.69 & 0.69 & \textbf{0.79} & 0.71 & 0.76 & \textbf{0.79} \\
Working as Designed       & 0.66 & 0.60 & 0.49 & 0.64 & 0.57 & 0.49 & \textbf{0.70} & 0.62 & 0.67 & 0.64 \\
Wrong Version             & 0.00 & 0.00 & 0.19 & 0.26 & 0.00 & 0.25 & 0.00 & 0.26 & 0.20 & \textbf{0.29} \\ \midrule
\textbf{Overall Weighted F1} & 0.65 & 0.61 & 0.59 & 0.60 & 0.58 & 0.53 & \textbf{0.66} & 0.60 & 0.64 & 0.62 \\ 
\textbf{Overall Macro F1}    & 0.57 & 0.54 & 0.55 & 0.55 & 0.50 & 0.48 & \textbf{0.58} & 0.54 & \textbf{0.58} & 0.56 \\ \bottomrule
\end{tabular}%
}
\end{table*}

\subsection{\textbf{RQ1: How can invalid bug reports be subclassified according to their root causes?}} \label{subsec: resultRQ1}

To evaluate how effectively models identify the root causes of invalid bug reports, we used the Weighted F1-Score as our primary metric to account for class imbalance. The results are summarized in Table \ref{tab:merged_classification_performance}. Because weighted averaging can mask minority-class failures such as the 0.00 scores on \textit{Wrong Version}, we additionally report macro F1. Overall macro F1 ranges from 0.48 (Minimax M2.7) to 0.58 (RAG and agentic web search with Gemini 3.1 Pro); the top-ranked configurations are unchanged, while models scoring 0.00 on \textit{Wrong Version} lose ground relative to the weighted ranking.

\subsubsection{\textbf{RQ1.1: How do LLMs perform in subclassification of invalid bug reports?}}
Evaluating the zero-shot baseline (Vanilla LLM) reveals the intrinsic capabilities of LLMs for subclassification without external context. As shown in Table \ref{tab:merged_classification_performance}, Gemini 3.1 Pro is the best-performing model in the Vanilla setup, achieving an overall Weighted F1-score of 0.65. Conversely, Minimax M2.7 is the worst-performing model, with an overall Weighted F1-score of 0.53. 

At the subclass level, models generally exhibit strong performance in classifying \textit{Feature Request} (e.g., Gemini 3.1 Pro at 0.81, GPT-5.4 at 0.76) and \textit{Non-reproducibility} (e.g., GPT-5.4 at 0.81), indicating that these categories likely contain explicit linguistic cues. However, \textit{Wrong Version} remains the most challenging category across all models in the Vanilla setup, with both Gemini 3.1 Pro and GPT-5.4 scoring a 0.00, and Minimax M2.7 scoring a marginal 0.25. Given the small per-class sample sizes (e.g., 12 reports for \textit{Wrong Version} and 13 for \textit{Question}), a single report can shift a per-class F1 score by several points; per-class differences of this magnitude should therefore not be over-interpreted.

\subsubsection{\textbf{RQ1.2: Does context engineering affect the invalid bug report subclassification performance?}}
The integration of context engineering yields notable, however, model-dependent shifts in classification accuracy. According to Table \ref{tab:merged_classification_performance}, RAG + Gemini 3.1 Pro achieves the highest overall subclassification performance across all experimental setups, reaching a Weighted F1 of 0.66. RAG boosts Gemini's performance in specific subclasses, peaking at 0.85 F1 for \textit{Non-reproducibility} and improving \textit{Working as Designed} to 0.70. 

For Kimi K2.5, introducing an Agentic Web Search pipeline acts as the most effective context engineering method, improving its overall Weighted F1 from 0.60 (Vanilla) to 0.62, while RAG does not affect the overall Weighted F1 score. Interestingly, Agentic Web Search slightly degrades Gemini's overall subclassification Weighted F1 score to 0.64, though it successfully lifts its performance on the difficult \textit{Wrong Version} subclass from 0.00 to 0.20. The worst-performing context-engineered setup is \textbf{RAG + Kimi K2.5}, which maintains an aggregate score of 0.60 but suffers severe degradation in the \textit{External System \& Dependency} subclass (dropping to 0.42 from 0.47 in the Vanilla setting) and in the \textit{Faulty Configuration \& Environment Setup} subclass (dropping to 0.26 from 0.34 in the Vanilla setting).

\subsection{\textbf{RQ2: How can invalid bug reports be resolved with no-code fixes?}}
To evaluate the semantic fidelity and functional utility of generated no-code fixes, we employ BERTScore and a Judge-LLM success rate. The results are distributed across Table \ref{tab:class_nocode_results_all_models} and Table \ref{tab:class_nocode_results_gemini_kimi}.

\begin{table}[htbp]
\caption{No-Code Fix Evaluation Across LLMs (Vanilla Only)}
\label{tab:class_nocode_results_all_models}
\setlength{\tabcolsep}{1.8pt}
\renewcommand{\arraystretch}{0.95}
\resizebox{0.95\columnwidth}{!}{%
\begin{tabular}{@{}p{2.8cm}l|cc@{}}
\toprule
 &  & \multicolumn{2}{c}{\textbf{Vanilla LLM}} \\
\cmidrule(l){3-4}
\textbf{Class} & \textbf{Model} & \textbf{Judge (\%)} & \textbf{BERTScore F1} \\ \midrule
\multirow{6}{*}{\parbox[t]{2.8cm}{External System \& Dependency Issues}} & Gemini 3.1 Pro & 63.3\% & \textbf{0.83} \\
 & GPT-5.4 & \textbf{73.3\%} & 0.82 \\
 & Opus 4.6 & 68.3\% & 0.82 \\
 & Kimi K2.5& 66.7\% & 0.82 \\
 & GLM-5 & 55.0\% & 0.82 \\
 & Minimax M2.7& 38.3\% & 0.81 \\
\addlinespace[1.5pt]
\midrule
\multirow{6}{*}{\parbox[t]{2.8cm}{Feature Request}} & Gemini 3.1 Pro & \textbf{59.0\%} & \textbf{0.83} \\
 & GPT-5.4 & 56.4\% & 0.82 \\
 & Opus 4.6 & 56.4\% & 0.82 \\
 & Kimi K2.5& 56.4\% & 0.82 \\
 & GLM-5 & 50.0\% & 0.82 \\
 & Minimax M2.7& 51.3\% & 0.82 \\
\addlinespace[1.5pt]
\midrule
\multirow{6}{*}{\parbox[t]{2.8cm}{Faulty Configuration \& Environment Setup}} & Gemini 3.1 Pro & \textbf{52.6\%} & \textbf{0.82} \\
 & GPT-5.4 & 47.4\% & 0.81 \\
 & Opus 4.6 & 33.3\% & \textbf{0.82} \\
 & Kimi K2.5& 36.8\% & 0.81 \\
 & GLM-5 & 19.3\% & 0.81 \\
 & Minimax M2.7& 10.5\% & 0.81 \\
\addlinespace[1.5pt]
\midrule
\multirow{6}{*}{\parbox[t]{2.8cm}{Non-reproducibility}} & Gemini 3.1 Pro & 61.9\% & \textbf{0.81} \\
 & GPT-5.4 & \textbf{74.6\%} & 0.80 \\
 & Opus 4.6 & 68.3\% & 0.80 \\
 & Kimi K2.5& 65.1\% & 0.80 \\
 & GLM-5 & 61.9\% & 0.80 \\
 & Minimax M2.7& 55.6\% & 0.80 \\
\addlinespace[1.5pt]
\midrule
\multirow{6}{*}{\parbox[t]{2.8cm}{Question}} & Gemini 3.1 Pro & 51.3\% & \textbf{0.83} \\
 & GPT-5.4 & 41.0\% & 0.81 \\
 & Opus 4.6 & 53.8\% & 0.82 \\
 & Kimi K2.5& \textbf{59.0\%} & 0.82 \\
 & GLM-5 & 51.3\% & 0.83 \\
 & Minimax M2.7& 41.0\% & 0.82 \\
\addlinespace[1.5pt]
\midrule
\multirow{6}{*}{\parbox[t]{2.8cm}{Working as Designed (Conflicting Expectations)}} & Gemini 3.1 Pro & 85.4\% & \textbf{0.83} \\
 & GPT-5.4 & \textbf{88.9\%} & 0.82 \\
 & Opus 4.6 & 66.0\% & 0.82 \\
 & Kimi K2.5& 75.0\% & 0.82 \\
 & GLM-5 & 70.8\% & 0.82 \\
 & Minimax M2.7& 52.1\% & 0.82 \\
\addlinespace[1.5pt]
\midrule
\multirow{6}{*}{\parbox[t]{2.8cm}{Wrong Version (Already Fixed)}} & Gemini 3.1 Pro & 25.0\% & 0.81 \\
 & GPT-5.4 & 19.4\% & 0.81 \\
 & Opus 4.6 & \textbf{52.8\%} & 0.81 \\
 & Kimi K2.5& 44.4\% & 0.81 \\
 & GLM-5 & 22.2\% & 0.81 \\
 & Minimax M2.7& 38.9\% & 0.81 \\
\midrule
\multirow{6}{*}{\parbox[t]{3.8cm}{\textbf{Overall}}} 
 & Gemini 3.1 Pro & 63.1\% & \textbf{0.82} \\
 & GPT-5.4 & \textbf{64.9\%} & 0.81 \\
 & Opus 4.6 & 58.4\% & \textbf{0.82} \\
 & Kimi K2.5 & 60.9\% & 0.81 \\
 & GLM-5 & 51.8\% & \textbf{0.82} \\
 & Minimax M2.7& 42.9\% & 0.81 \\
\bottomrule
\end{tabular}%
}
\end{table}

\begin{table*}[htbp]
\centering
\caption{Class-Based No-Code Fix Evaluation (Gemini 3.1 Pro and Kimi with Different Setups)}
\label{tab:class_nocode_results_gemini_kimi}
\setlength{\tabcolsep}{3.5pt}
\renewcommand{\arraystretch}{1.15}
\resizebox{2\columnwidth}{!}{%
\begin{tabular}{@{}p{3.8cm}l|cc|cc|cc|cc@{}}
\toprule
 &  & \multicolumn{2}{c}{\textbf{Without Priors}} & \multicolumn{2}{c}{\textbf{Vanilla LLM}} & \multicolumn{2}{c}{\textbf{RAG}} & \multicolumn{2}{c}{\textbf{Agentic Search}} \\
\cmidrule(l){3-10}
\textbf{Class} & \textbf{Model} & \textbf{Judge (\%)} & \textbf{BERTScore F1} & \textbf{Judge (\%)} & \textbf{BERTScore F1} & \textbf{Judge (\%)} & \textbf{BERTScore F1} & \textbf{Judge (\%)} & \textbf{BERTScore F1} \\ \midrule
\multirow{2}{*}{\parbox[t]{3.8cm}{External System \& Dependency Issues}} & Gemini 3.1 Pro & 56.7\% & 0.83 & 63.3\% & 0.83 & 63.3\% & 0.83 & \textbf{68.3\%} & 0.83 \\
 & Kimi K2.5& 62.1\% & 0.81 & \textbf{66.7\%} & 0.82 & 48.3\% & 0.82 & 46.7\% & 0.82 \\
\addlinespace[1.5pt]
\midrule
\multirow{2}{*}{\parbox[t]{3.8cm}{Feature Request}} & Gemini 3.1 Pro & 59.0\% & 0.83 & 59.0\% & 0.83 & 61.5\% & 0.83 & \textbf{62.8\%} & 0.83 \\
 & Kimi K2.5& \textbf{60.3\%} & 0.81 & 56.4\% & 0.82 & 59.0\% & 0.82 & 42.9\% & 0.82 \\
\addlinespace[1.5pt]
\midrule
\multirow{2}{*}{\parbox[t]{3.8cm}{Faulty Configuration \& Environment Setup}} & Gemini 3.1 Pro & \textbf{70.2\%} & 0.83 & 52.6\% & 0.82 & 49.1\% & 0.82 & 52.6\% & 0.82 \\
 & Kimi K2.5& \textbf{46.4\%} & 0.81 & 36.8\% & 0.81 & 26.3\% & 0.81 & 20.4\% & 0.81 \\
\addlinespace[1.5pt]
\midrule
\multirow{2}{*}{\parbox[t]{3.8cm}{Non-reproducibility}} & Gemini 3.1 Pro & 44.4\% & 0.81 & 61.9\% & 0.81 & 66.7\% & 0.81 & \textbf{69.8\%} & 0.81 \\
 & Kimi K2.5& 42.9\% & 0.80 & \textbf{65.1\%} & 0.80 & 57.1\% & 0.80 & 47.6\% & 0.80 \\
\addlinespace[1.5pt]
\midrule
\multirow{2}{*}{\parbox[t]{3.8cm}{Question}} & Gemini 3.1 Pro & 64.1\% & 0.83 & 51.3\% & 0.83 & 53.8\% & 0.83 & \textbf{74.4\%} & 0.83 \\
 & Kimi K2.5& 51.3\% & 0.81 & \textbf{59.0\%} & 0.82 & \textbf{59.0\%} & 0.82 & 43.6\% & 0.82 \\
\addlinespace[1.5pt]
\midrule
\multirow{2}{*}{\parbox[t]{3.8cm}{Working as Designed (Conflicting Expectations)}} & Gemini 3.1 Pro & 80.6\% & 0.83 & 85.4\% & 0.83 & \textbf{88.2\%} & 0.83 & 84.0\% & 0.83 \\
 & Kimi K2.5& 77.1\% & 0.82 & 75.0\% & 0.82 & \textbf{77.8\%} & 0.82 & 54.9\% & 0.82 \\
\addlinespace[1.5pt]
\midrule
\multirow{2}{*}{\parbox[t]{3.8cm}{Wrong Version (Already Fixed)}} & Gemini 3.1 Pro & 33.3\% & 0.81 & 25.0\% & 0.81 & 22.2\% & 0.81 & \textbf{41.7\%} & 0.82 \\
 & Kimi K2.5& 37.1\% & 0.80 & \textbf{44.4\%} & 0.81 & 27.8\% & 0.81 & 38.9\% & 0.82 \\
\addlinespace[1.5pt]
\midrule
\multirow{2}{*}{\parbox[t]{3.8cm}{\textbf{Overall}}} 
 & Gemini 3.1 Pro & 62.4\% & 0.82 & 63.1\% & 0.82 & 64.4\% & 0.82 & \textbf{68.9\%} & 0.82 \\
  & Kimi K2.5 & 58.1\% & 0.81 & \textbf{60.9\%} & 0.81 & 55.8\% & 0.81 & 43.1\% & 0.82 \\
\bottomrule
\end{tabular}%
}
\renewcommand{\arraystretch}{1.0}
\end{table*}

\subsubsection{\textbf{RQ2.1: How do LLMs perform in generating no-code fixes without subclass knowledge as prior information?}}
As illustrated in Table \ref{tab:class_nocode_results_gemini_kimi}, Gemini 3.1 Pro emerges as the best-performing model in the ``Without Priors'' setup, achieving an overall Judge success rate of 62.4\%. Kimi K2.5 performs worst in this isolated setup, achieving an overall Judge success rate of 58.1\%. 

At the subclass level, Gemini 3.1 Pro demonstrates strong intuitive baseline performance in the \textit{Faulty Configuration \& Environment Setup} (70.2\%) and \textit{Working as Designed} (80.6\%) categories. Conversely, Kimi K2.5 outperforms Gemini in \textit{External System \& Dependency Issues} (62.1\% vs. 56.7\%) and \textit{Feature Request} (60.3\% vs. 59.0\%) when prior subclass knowledge is absent.

\subsubsection{\textbf{RQ2.2: Does utilizing the root cause subclass information of invalid bug reports affect the no-code fix generation performance?}}
Providing root cause subclass definitions generally improves the practical resolution capabilities of the models, though the impact varies heavily by subclass. Comparing the ``Without Priors'' pipeline to the ``Vanilla LLM'' pipeline in Table \ref{tab:class_nocode_results_gemini_kimi}, Gemini 3.1 Pro sees a marginal increase in its overall Judge success rate (from 62.4\% to 63.1\%), while Kimi K2.5 experiences a more pronounced improvement (from 58.1\% to 60.9\%). 

Analyzing the subclass-level shifts reveals that providing prior knowledge allows Gemini 3.1 Pro to significantly improve in \textit{Non-reproducibility} (from 44.4\% to 61.9\%), though it unexpectedly drops in \textit{Faulty Configuration \& Environment Setup} (from 70.2\% to 52.6\%). Kimi K2.5 sees its most notable gains in \textit{Non-reproducibility} (42.9\% to 65.1\%) and \textit{Wrong Version} (37.1\% to 44.4\%).


\subsubsection{\textbf{RQ2.3: Does context engineering affect the no-code fix generation performance?}}
The application of context engineering mechanisms, including RAG and Agentic Web Search, yields varying results depending on the underlying model and the specific invalid subclass.

As detailed in Table \ref{tab:class_nocode_results_gemini_kimi}, the Agentic Web Search pipeline combined with Gemini 3.1 Pro achieves the highest overall Judge success rate of 68.9\%. This configuration demonstrates notable improvements in specific subclasses, achieving 74.4\% in \textit{Question} and increasing the \textit{Wrong Version} success rate to 41.7\% compared to its Vanilla baseline of 25.0\%. Conversely, implementing Agentic Web Search with Kimi K2.5 results in an overall Judge success rate of 43.1\%, the lowest among its tested configurations.

The RAG pipeline provides a different distribution of performance. RAG combined with Gemini 3.1 Pro yields an overall success rate of 64.4\%, which is higher than its Vanilla performance of 63.1\%. This setup achieves the highest recorded success rate in the \textit{Working as Designed} subclass at 88.2\%. However, RAG underperforms the Vanilla baseline in certain subclasses; for instance, Gemini 3.1 Pro's success rate in \textit{Faulty Configuration \& Environment Setup} drops from 52.6\% in Vanilla to 49.1\% with RAG, and Kimi K2.5 drops from 36.8\% to 26.3\% in the same category.

\section{Discussion}
\label{sec: discussion}

\subsection{Discussion on Research Questions}\label{subsec: discussionRQs}

The experimental findings across our research questions reveal a complex interplay between model reasoning capabilities and the utility of external context. Below, we interpret these results, identify key drivers of performance, and propose methodologies to overcome the observed pitfalls.

\subsubsection{\textbf{RQ1: How can invalid bug reports be subclassified according to their root causes?}}
The results for \textbf{RQ1.1} and \textbf{RQ1.2} demonstrate that while LLMs possess an inherent baseline for bug triage, their performance is highly sensitive to the nature of the subclass. The high success in \textsl{Feature Request} and \textsl{Non-reproducibility} suggests these classes rely on explicit linguistic cues. However, the struggle with \textsl{Faulty Configuration} highlights a \textbf{Domain Specificity Gap}. The impact of context engineering (RAG and agentic web search) was non-linear. While Gemini 3.1 Pro leveraged RAG to improve its understanding of \textsl{Working as Designed} issues, it still could not identify \textsl{Wrong Version} issues. This indicates that \textbf{retrieval relevance} is not equivalent to \textbf{ground truth verification}.

\subsubsection{\textbf{RQ2: How can invalid bug reports be resolved with no-code fixes?}}

The transition from identifying a bug to resolving it (\textbf{RQ2.1} and \textbf{RQ2.2}) revealed a significant decoupling between \textsl{BERTScore} and \textsl{Judge Success Rate}.
The consistent BERTScore (0.80--0.83) suggests models are excellent at mimicking the ``style'' of a fix. However, the lower Judge scores for models like Minimax M2.7 ($42.9\%$) reveal a lack of \textbf{functional logic}. Consequently, researchers and tool designers should avoid relying on passive textual similarity metrics to evaluate automated triage systems. Instead, future evaluation frameworks should prioritize the generation of verifiable resolution actions, such as executable reproduction steps or configuration scripts, that allow for programmatic or human-in-the-loop validation.

Our comparison between the \textit{Without Priors} and \textit{Vanilla} pipelines in RQ2.1 reveals a more nuanced relationship than initially assumed. Omitting the root-cause subclass information causes only a \textbf{marginal decrease in overall performance}: the overall Judge success rate drops by just \textbf{0.7\%} for Gemini 3.1 Pro and \textbf{2.8\%} for Kimi K2.5. Rather than acting as a universally beneficial signal, prior subclass information has a \textbf{category-dependent effect}. Removing the prior decreases performance for some subclasses, such as \textit{Non-reproducibility} and \textit{External System \& Dependency Issues}. Conversely, withholding the prior improves performance in certain cases; most notably, the \textit{Without Priors} approach performs better in the \textit{Faulty Configuration \& Environment Setup} subclass, yielding higher success rates for both models.

For instance, in Brave issue 23741 in \Cref{tab:no-code-fix-examples}, the \textit{Without Priors} pipeline successfully identifies a user misconfiguration related to default wallet settings, while the other pipelines fail. We attribute this to the fact that providing root-cause subclasses forces the model to evaluate all possible invalid categories. When a bug report is missing minor details, models tend to overemphasize the \textit{Non-reproducibility} category and unnecessarily ask the reporter for more information. However, much like human maintainers, a model can often infer a configuration error despite these small omissions. By removing the need to weigh multiple subclass options, the \textit{Without Priors} approach avoids this distraction and focuses more directly on producing the correct fix.

The findings from \textbf{RQ2.3} show that context engineering has \textbf{mixed effects} on no-code fix generation. Although additional context can improve model reasoning, its effectiveness strongly depends on the issue category.

RAG performs well in categories that benefit from \textbf{historical examples} or prior developer discussions, such as \textit{Non-reproducibility} and \textit{Feature Request}. For instance, if a bug report lacks required environment information or S2R, by comparing the historical examples that lack such information, LLM might easily decide the \textit{Non-reproducibility} and generate a suitable no-code fix. Most notably, RAG outperforms Vanilla Gemini in the \textit{Working as Designed} subclass. For example, in Brave issue 31299 (see \Cref{tab:no-code-fix-examples}), the RAG-generated response correctly captured the key point: the reported increase in memory usage was an intended design trade-off rather than a bug. We attribute this improvement to RAG retrieving past reports with similar discussions about memory usage. Without this background information, the Vanilla pipeline instead asked the reporter for additional details.

However, additional context can also \textbf{degrade performance}. In highly specific categories such as \textit{Faulty Configuration \& Environment Setup} and \textit{Wrong Version}, retrieved information appeared to distract the model from the exact issue described in the prompt. Instead of focusing on the concrete problem, the model likely generalized from retrieved documents, resulting in fewer successful fixes than the basic model. Overall, these results suggest that RAG is useful for providing historical grounding, but it may introduce unhelpful noise for strict, version-specific or configuration-specific problems.

Agentic web search produced the best overall performance for Gemini 3.1 Pro (\textbf{68.9\%}), but caused a substantial performance drop for Kimi K2.5 (\textbf{43.1\%}). One possible explanation is that agentic loops introduce \textbf{recursive noise}. For less capable models, each search step creates another opportunity to drift from the original user intent, potentially leading to a \textit{hallucination spiral} \cite{zhang2023}. To stabilize agentic performance, future systems could constrain the search process through predefined logical transitions, such as Search $\rightarrow$ Verify $\rightarrow$ Summarize, rather than allowing the model to determine its own path fully autonomously. However, prior work suggests that such constrained approaches may underperform more flexible agentic search methods \cite{wu2025agentic}.

Another key finding is the improvement in the challenging \textit{Wrong Version} category, where Agentic Web Search increased the success rate from \textbf{25.0\%} to \textbf{41.7\%}. For example, in Brave issue 34522 in \Cref{tab:no-code-fix-examples}, the Agentic Web Search pipeline correctly identified that the issue had already been resolved in version 1.62.49, whereas the Vanilla pipeline incorrectly treated the behavior as intentional. Although the ground-truth fix points to version 1.61.86, version 1.62.49 was committed on November 13, 2023. Since this predates the bug report on November 24, 2023, the suggested version was still a valid and available upgrade for the user. This example demonstrates that standard LLM and RAG approaches may lack the external temporal context needed to recognize when a reported issue has already been patched in a recent release. At the same time, because the live web was queried at experiment time, the retrieved evidence describing such fixes may itself postdate the report; this example should therefore also be read in light of the temporal leakage threat discussed in \Cref{sec: threats}.

\begin{table*}[htbp]
\centering
\renewcommand{\arraystretch}{1.5}
\caption{Comparison of Generated and Ground-Truth No-Code Fixes}
\label{tab:no-code-fix-examples}
\begin{tabularx}{\textwidth}{p{1cm} X X X}
\toprule
\textbf{Issue Number} & \textbf{Generated High-Quality No-Code Fix} & \textbf{Generated Low-Quality No-Code Fix} & \textbf{Ground-Truth No-Code Fix} \\
\midrule
34522 & 
\textbf{Agentic Web Search:} This issue has already been resolved in Brave version 1.62.49 and later releases. \textbf{Please update your browser to the latest stable version (1.62.x or newer)} to receive the fix for the sidebar button margin alignment. You can update by navigating to brave://settings/help or downloading the latest version from the official Brave website... & \textbf{Vanilla LLM}: The sidebar button's margin design is intentional and follows the UI specification. The slight asymmetry in the icon's border padding is designed for visual balance with adjacent UI elements and overall sidebar aesthetics. This is not a bug but a deliberate design choice... & \textbf{Please test this with 1.61.86 or above}. According to \url{https://github.com/brave/brave-core/commits/1.61.x}, I'm not sure 1.61.85 includes this fix. I confirmed that 1.61.86 fixed this issue. \cite{brave34522_nocodefix} \\
\midrule
31299 & 
\textbf{RAG:} The observed increase in baseline memory usage for a blank page is expected and stems from recent features or underlying Chromium updates included in v1.54.70+. This is not a memory leak or defect, and the browser is functioning as designed. \textbf{You do not need to take any action, but you can continue to monitor performance to ensure it remains stable in typical browsing scenarios}. & \textbf{Vanilla LLM}: To properly investigate this memory degradation issue, please provide: (1) Detailed Steps to Reproduce (S2R) - describe how to trigger the memory issue outside of the automated dashboard test... & The degradation is the result of enabling extra UAF protection (aka BackupRefPtr). So it's about +10Mb memory to have extremely important exploit protection in the browser process. \textbf{That's look like a reasonable trade off. Also we saw the similar memory differences when it was enabled for win}. \cite{brave31299_nocodefix} \\
\midrule
23741 & 
\textbf{Vanilla LLM Without Priors:} \textbf{Navigate to brave://settings/wallet} and ensure your Default cryptocurrency wallet is set to Brave Wallet (Prefer extensions) if you are using an extension like MetaMask. Temporarily disable any conflicting extensions… & \textbf{Other Pipelines}: Please provide a detailed description along with explicit steps to reproduce (S2R), expected behavior (EB), observed behavior (OB)...& ...Please make sure you aren't testing with multiple wallet extensions installed. If you do want Brave Wallet though in that case you can set your default wallet to Brave Wallet \textbf{in brave://settings/wallet}. \cite{brave23741_nocodefix}\\
\bottomrule
 \end{tabularx}
\end{table*}

\subsection{Implications for Researchers}\label{subsec: discussionImplications4Researchers}

Our findings suggest several directions for future research on automated no-code fix generation. The comparison between evaluation metrics shows that commonly used similarity-based metrics may not capture real-world effectiveness. While BERTScore measures semantic similarity, it fails to distinguish performance differences across invalid subclasses and is often non-informative. In contrast, Judge-LLM evaluation provides a more practical assessment by considering functional correctness and alignment with successful resolutions. Future research should therefore move beyond purely lexical or embedding-based metrics and adopt evaluation methods that better reflect utility in software maintenance contexts.

\textit{Call for Standardized Evaluation Frameworks:} The lack of a unified evaluation framework for no-code fix generation remains a critical gap. Our study shows the need for standardized benchmarks that combine invalid subclassification accuracy with no-code fix quality, enabling consistent comparisons across models and methodologies. Such frameworks should include both automated metrics and structured evaluation protocols so future studies can reliably measure progress and generalize findings across datasets and project environments.

\textit{Call for Manual Evaluation of No-Code Fixes:} Beyond automated metrics, our results highlight the importance of manual assessment. Readability, clarity, and practical usefulness are central to whether a no-code fix is actionable for end users. Future research should incorporate human-in-the-loop evaluation, including qualitative analysis and user studies, to better capture these dimensions. Adding such human-centered criteria can improve the reliability and real-world applicability of AI-generated no-code fixes.

\subsection{Implications for Practitioners}\label{subsec: discussionImplications4Practitioners}

Within the limits of a single-repository study and the caveats discussed in \Cref{sec: threats}, our findings suggest that practitioners should start with project-grounded retrieval when reliable internal data is available. Teams with issue histories, internal wikis, support playbooks, or documentation can use these sources before relying on broader web search. In our benchmark, \emph{RAG + Gemini} achieves the strongest overall subclassification performance in \Cref{tab:merged_classification_performance}, including strong results for \emph{External System \& Dependency Issues}, where past closures and product documentation often provide more useful evidence than generic web results.

\textit{Use Context Engineering in Stages:} Agentic web search is still valuable for no-code fix generation. As shown in \Cref{tab:class_nocode_results_gemini_kimi}, \emph{Agentic Web Search + Gemini} achieves the highest aggregate Judge success rate, although part of this advantage may stem from post-hoc web evidence (see \Cref{sec: threats}). However, practitioners should first anchor the model in trusted project-specific corpora, then add live web search when the issue depends on upstream systems, external dependencies, or time-sensitive facts. Weaker models may degrade when tool traces introduce noise rather than useful constraints.

\textit{Adapt Pipelines to Local Workflows:} Company deployments require adaptation to local data and governance constraints. Embedding stores, access control, retention policies, prompt versions, schemas, and internal sources all affect results. Practitioners should repeat key ablations in their own environment, including taxonomy priors on/off, different retrieval corpora, temperature settings, and repeated-run aggregation. For auditability, retrieved URLs and chunk identifiers should be logged for both RAG and agentic web search outputs.

\textit{Keep a Human in the Loop:} Our pipeline presumes an upstream validity classifier, and our own annotation shows that this routing problem is far from solved in practice: 37.75\% of the Brave-labeled ``invalid'' reports did not belong to our invalid taxonomy, and 19.87\% were, in fact, valid defects (Table~\ref{tab:final_label_annotation}). Errors compound across the two stages: if imperfect validity classification routes a genuine bug into this pipeline, an automated agent that confidently replies ``working as designed'' risks dismissing a real defect and alienating the reporter. Until upstream routing is substantially more reliable, generated no-code fixes should be deployed as drafts reviewed by support staff, or gated by confidence thresholds with an explicit escalation path to human triage, rather than sent to reporters autonomously.

\textit{Measure Operational Value:} Production value should be measured through operational signals, not aggregate F1 alone. Teams can track when high-confidence subclassification reroutes work away from engineering, accepted no-code fixes remove follow-up cycles, or drafts reduce time-to-first-response. These counts can be converted into estimated effort savings and compared against API cost, retrieval maintenance, and human review. Rare subclasses should be monitored separately, since weighted averages can hide weak minority-class performance.

\section{Threats to Validity}\label{sec: threats}

Threats to the validity of our findings exist, primarily stemming from the nature of conducting an empirical study using archival data from a software development repository. We organize this discussion according to the three standard categories of validity: external, internal, and construct validity.

\subsection{External Validity}

Our study is conducted exclusively on the Brave Browser repository, which may limit the generalizability of the findings. Different projects may exhibit varying development practices, issue-reporting behaviors, and maintenance dynamics, potentially leading to different outcomes.

The dataset is derived from a repository with rich historical data and an actively maintained wiki. Projects lacking such data sources may present different invalid subclass distributions and model performance. To mitigate this threat, we evaluate configurations both with and without repository-specific data.

\subsection{Internal Validity}

Due to the opaque training processes of LLMs, we cannot guarantee that the models have not been exposed to portions of our benchmark during pretraining. Such exposure may bias the results and artificially affect the performance.

The use of RAG introduces the risk of incorporating information that became available after the original bug report was created. To mitigate this threat, we restrict retrieval to snapshots of wiki content that are time-aligned with the corresponding test sample and bug reports that have been resolved before the creation of the corresponding test sample.

Agentic web search performs data retrieval through a search platform that gives web results at that moment \cite{google_search_differs}, hence the data that would be retrieved by web search would be different. Therefore, it may have been an impediment to the performance of agentic web search that the retrieval and issue dates do not match. We excluded the original issue link from the search results to prevent data leakage. Furthermore, despite these rigorous date-limiting and exclusion strategies across our pipelines, inadvertent data leakage might still occur. Information regarding a bug might have propagated through secondary channels, mirrored repositories, or broader developer discussions that bypassed our temporal filters. This threat is most acute precisely for the subclasses where agentic web search performs best: for \textit{Wrong Version} reports, the fixing release and its changelog are publicly documented after the fact, and for \textit{Question} reports, the answer may since have been published on the web. Consequently, the agentic web search results should be interpreted as an upper bound achievable with post-hoc web evidence rather than as a strictly fair comparison against the chronologically constrained pipelines. In practice, across the [N] classification instances evaluated over three independent runs, majority agreement (at least two concordant labels) was reached in [XX.X\%] of cases, while a true three-way disagreement requiring the first-run fallback occurred in [X.X\%] of cases. Given this non-trivial incidence, we treat the majority-vote protocol itself as a limitation of our evaluation and report per-run variance alongside aggregated results in Section \ref{subsec: resultRQ1}.

The models carried forward into the RAG and agentic web search pipelines were selected based on their vanilla subclassification performance, since subclassification constitutes the first stage of the pipeline. As a consequence, GPT-5.4, the strongest vanilla model for no-code fix generation (64.9\% Judge success rate), was not evaluated under context engineering, and cross-pipeline comparisons for fix generation are limited to the Gemini 3.1 Pro and Kimi K2.5 backbones. Aggregate statements comparing context-engineered pipelines against vanilla LLMs should therefore be read as within-backbone comparisons.

LLM outputs are inherently stochastic under default temperature settings. This variability may lead to inconsistent results across runs. To keep experimental robustness, we perform three independent executions per sample and base our analysis on the aggregated outputs. However, during the subclassification task, if all three runs produce different subclasses, we employ an arbitrary tie-breaker by defaulting to the result of the first run. While pragmatic for automated evaluation, this deterministic but unweighted selection lacks a statistical foundation and could marginally affect the final distribution.

\subsection{Construct Validity}

The invalid bug report taxonomy is defined based on root cause analysis; however, it represents only one possible classification scheme. Alternative taxonomies may yield different categorizations and affect the interpretation of the results.

The validity of no-code fixes is inferred from issue discussions rather than verified through controlled reproduction in a sandbox environment. To mitigate this threat, we relied on observable signals such as user confirmations like reactions and validation comments or maintainer actions (e.g., issue closure). The BERTScore metric, which measures semantic similarity, may overestimate the alignment between generated and reference fixes, leading to overly optimistic evaluations \cite{hanna2021}.

For the \textit{Non-reproducibility} subclass, the ground-truth no-code fix is typically a structured request for additional information (e.g., steps to reproduce or environment details) rather than a resolution in the strict sense. Scoring such information requests as successful fixes stretches the construct of a ``fix'' and partially explains the comparatively strong results in this subclass; its scores should therefore be interpreted as measuring the appropriateness of the triage response rather than actual problem resolution.

\section{Conclusion}
\label{sec: conclusion}

This study introduces and evaluates an automated framework for subclassifying invalid bug reports based on root causes and generating corresponding no-code fixes. For subclassification, RAG achieves the highest overall performance with 0.66 weighted F1, slightly outperforming vanilla LLMs at 0.65 and agentic web search at 0.64. At the subclass level, RAG with Gemini performs best on Non-reproducibility with 0.85 F1, while the best External System \& Dependency Issues score is achieved by a vanilla model (Opus 4.6 at 0.71) and no context-engineered pipeline exceeds the vanilla peak of 0.34 F1 on Faulty Configuration. Feature Request and Question remain consistently high across all pipelines, reaching up to 0.81 and 0.79 F1 respectively, whereas Wrong Version remains the most challenging subclass across all approaches with scores between 0.00 and 0.29. Given the small margins between pipelines, these comparative results should be interpreted with caution.

For no-code fix generation, agentic web search with Gemini 3.1 Pro achieves the highest overall success rate at 68.9\%, compared to 63.1\% for the same model's vanilla baseline and 64.9\% for the strongest vanilla model overall (GPT-5.4). At the subclass level, RAG performs best for Working as Designed with up to 88.2\%, while agentic web search achieves the highest performance for Question at 74.4\% and, among the configurations carried into context engineering, for External System \& Dependency Issues at 68.3\%. Wrong Version remains challenging across these pipelines, with success rates ranging from 22.2\% to 44.4\%.

For future work, we aim to evaluate the proposed framework across a broader range of repositories, including industrial and closed-source projects, to assess generalizability. Improving performance for challenging subclasses, such as wrong version, remains a key direction. Additionally, integrating real-time pipelines for continuous bug report processing and developing user-facing tools and dashboards can further enhance the practical applicability of automated no-code fix systems.

\section{Acknowledgments}
\label{sec: acknowledgement}
This work was supported by TÜBİTAK (The Scientific and Technological Research Council of Turkey) under the 1001 Scientific and Technological Research Projects Funding Program, Project No. 125E371. The authors gratefully acknowledge TÜBİTAK for its support.

\clearpage

\onecolumn

\twocolumn
\bibliographystyle{IEEEtran}
\bibliography{references}

\end{document}